\documentclass[a4paper,10pt,nofootinbib]{article}

\usepackage[margin=2.84cm]{geometry}
\usepackage{epsf,epsfig,subfigure,graphicx,amsmath,amssymb,array}
\usepackage[sort&compress,numbers]{natbib}
\usepackage[svgnames]{xcolor}
\usepackage[colorlinks=true,urlcolor=MediumBlue,linkcolor=Crimson,citecolor=Crimson]{hyperref}
\usepackage[normalem]{ulem}
\usepackage{multirow}
\usepackage{makecell}

\usepackage{comment}
\usepackage{xcolor}
\usepackage{ulem}
\usepackage{cleveref}
\usepackage{url}

\def\sw0{{$\sin^2\theta_W^0$}}

\title{\bf Axion Dark Matter: How to see it?}

\author{\bf Yannis K. Semertzidis$^{1,2}$ and SungWoo Youn$^{1}$}

\begin{document}
\maketitle

\begin{abstract} 
The axion is a highly motivated elementary particle which could address two fundamental questions in physics - the strong CP problem and the dark matter mystery.
Experimental searches for this hypothetical particle have started to reach theoretically interesting sensitivity levels, particularly in the $\mu$eV (GHz) region.
They rely on large volume solenoid magnetic fields and microwave resonators with signals read out by quantum noise limited amplifiers.
Concurrently, there have been intensive experimental efforts to widen the search range by devising various techniques as well as to enhance sensitivities by implementing advanced technologies.
The developments and improvements in these orthogonal approaches will enable us to explore most of the parameter space of the axion and axion-like particles within the next couple of decades, with the 1-25 GHz frequency range to be conquered well within the first decade.
We review the experimental aspects of axion physics and discuss the past, present and future of the direct search programs.
\\\\
{\it A review prepared for Science Advances}\\
\\
$^{1}$ Center for Axion and Precision Physics Research, IBS, Daejeon 34051, Republic of Korea\\
$^{2}$ Department of Physics, KAIST, Daejeon 34141, Republic of Korea\\

%\keywords{``Invisible'' axion; GHz range axions; High field magnet.}
\end{abstract}
%\pacs{,,}

%%%%%%%%%%%%%%%%%%%%%%%%%%%%%%%%%%%%%%%%%%%%%%%%%%%%%%%%%%%%%%%%%%%%%%%%%%%%%%%%%%%%%%%%%%%%%%
\section{Introduction}\label{sec:introduction}
As discussed in the companion theoretical review, the Peccei-Quinn (PQ) mechanism offers a dynamic solution to the $CP$ symmetry problem in quantum chromodynamics (QCD) of particle physics.
The mechanism involves a new global U(1) symmetry with an associated pseudoscalar field permeating all space. 
The symmetry is assumed to have been spontaneously broken at a certain energy scale while the Universe was evolving at its early stages.
This spontaneous process of symmetry breaking is supposed to be accompanied by the production of a pseudo-Nambu-Goldstone boson, the axion~\cite{article:PQ77, article:Weinberg78, article:Wilczek78}.
The null results from searches for the standard axion with a mass of  order of a few hundred keV have quickly diverted attention to ``invisible" axions with very small masses~\cite{article:KSVZ1}. 
The QCD axion has theoretically well-defined properties: 1) its mass determined by the symmetry-breaking energy scale; and 2) its interactions to Standard Model particles depending on models, Kim-Shifman-Vainshtein-Zakharov (KSVZ)~\cite{article:KSVZ1, article:KSVZ2} and Dine-Fischler-Srednicki-Zhitnitsky (DFSZ)~\cite{article:DFSZ1, article:DFSZ2}.
There is a more generic type of axions, called the axion-like particle (ALP), which is not directly related to QCD but favored by certain theoretical models including string theory~\cite{article:string1, article:string2, article:string3}.
The QCD axions and ALPs, together referred to as axions, could also account for the halos of dark matter, the mysterious substance that is believed to constitute $\sim 85\%$ of the matter in the Universe~\cite{article:axion_cosmology1, article:axion_cosmology2, article:axion_cosmology3}.

Not yet revealed, the axion models have been extensively studied over last 40 years with growing attention and tremendous efforts have been made to test them experimentally. 
Several experimental searches are presently underway mainly exploiting the axion field interaction with the electromagnetic fields.
The emergence of electromagnetic radiation, i.e., photons, out of vacuum in the presence of magnetic fields can be a distinctive signature of the axion in the galactic dark matter halo floating around us.
Direct conversion searches require a strong magnetic field to have a chance with the weak axion signal.
The Axion Dark Matter eXperiment (ADMX) at the University of Washington uses a resonant cavity immersed in a superconducting (SC) magnet to detect the weak conversion of axions to microwave photons.
Axions or axion-like particles produced by astrophysical objects could be detected by terrestrial telescopes.
The CERN Axion Search Telescope (CAST) was designed to detect axions produced in the Sun's core where X-rays scatter off strong electric fields.
Axions could also resonantly convert into photons in the magnetosphere of neutron stars which can leave distinct features in the spectra detectable by current space or ground telescopes.
Another technique exploits the photon-axion-photon oscillation to produce and detect the axion signal in the laboratory using a strong light source and a pair of magnets.
The Axion-Like Particle Search (ALPS) shines a laser beam into the vacuum on one side of a magnet setup to convert the light into axions, which pass through an optical wall to enter the other side, where they are converted back into light to shine out of vacuum.

\begin{figure}
    \centering
    \includegraphics[width=0.9\linewidth]{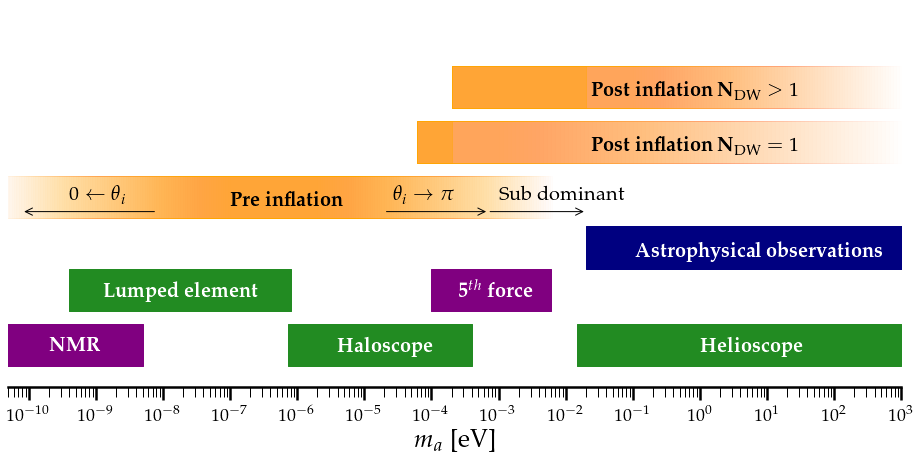}
    \caption{Allowed mass ranges for dark matter axions by various cosmological scenarios (in orange) and astrophysical observations (in navy).
    Different experimental approaches are involved to cover different mass ranges depending on the axion couplings (e.g., to photons in green or to nuclear spins in purple).
    }
    \label{fig:axion_mass}
\end{figure}

The main challenge of discovering the axion is that the particle mass is not theoretically predictable, but has to be experimentally determined by scanning a vastly wide energy range.
Some constraints on mass of axions as dark matter can be made by cosmological arguments and astrophysical observations~\cite{article:SN1987A}.
Despite such constraints, the search range spans several orders of magnitude between $\mu$eV and eV, with the lower bound possibly extended down to neV in certain cosmological models (see the accompanying theory section for more details).
In addition, the axion signal is expected to be very narrow with an equivalent quality factor of $Q\approx10^6$~\cite{article:Turner}, and extremely feeble due to very weak couplings to the Standard Model particles and fields. 
Recalling Eq.~(5) in the theory section, the interactions of the axion field, $a$, with the Standard Model fields are categorized into three types proportional to
\begin{equation}
   g_{a \gamma \gamma} \, a \, {\bf E} \cdot {\bf B},\,\,  g_{aff} \nabla a \cdot \hat{{\bf S}}, \,\,{\rm and}\,\, g_{\rm EDM} \, a \, \hat{{\bf S}} \cdot {\bf E},
\end{equation}
respectively, for electromagnetic fields, fermion spins, and nuclear electric dipole moments (EDMs).
The first type of interaction is examined with cavity resonators in strong magnetic fields (RBF~\cite{article:RBF1}, ADMX~\cite{article:ADMX}, HAYSTAC~\cite{article:HAYSTAC1}, CAPP~\cite{article:Yannis19}, etc.), the second interaction with magnetometry (GNOME~\cite{article:GNOME}, QUAX~\cite{proc:Caspers}, ARIADNE~\cite{article:ARIADNE}, etc.) and storage ring EDM methods~\cite{article:Graham21}, and the third one with nuclear magnetic resonance (NMR) spectroscopy (CASPEr~\cite{article:CASPEr}) and storage ring EDM methods~\cite{article:Chang19}. 
The probable axion mass range under various cosmological scenarios and the search regions by ongoing and upcoming experiments is given in Fig.~\ref{fig:axion_mass}.

Since the search range spans over several decades, it is inevitable that the technologies involved are also very different depending on the targeted mass range. 
In broad terms, the range of $10^{-6}-10^{-3}$\,eV involves the use of microwave resonators in a solenoid magnetic field and/or multi-layer dielectric resonators in a dipole magnetic field~\cite{article:Brun19}.
A higher mass range up to $\sim10^{-2}$\,eV can be covered by experiments sensitive to monopole-dipole interactions~\cite{article:new_force}, similar to techniques used in experiments looking for deviations from Newtonian mechanics at micrometer scales~\cite{article:non_Newtonian_force}.  
The search can be further extended above $\sim10^{-2}$\,eV by looking for solar axions, complementary to the astrophysical observations.
Below $\sim10^{-6}$\,eV, lumped element circuits with toroidal/solenoidal magnetic fields and NMR techniques are involved.
Experiments looking for static or oscillating nuclear EDM, e.g., neutron and proton EDM can also provide independent information on the relation between the axion field and the $\Bar\theta$ parameter of the Standard Model in low mass regions.

This article reviews major experimental endeavors to unveil the axion physics, while offering shorter descriptions for other efforts. 
Section~\ref{sec:strategies} introduces the detection principle and search strategies, and gives an overview of direct searches.
Through Secs.~\ref{sec:haloscope}$-$\ref{sec:LSW}, the individual searches will be described in more detail including a brief history and key features of the experimental research as well as their present status and future prospects.
A critical remark is made in Sec.~\ref{sec:axion_EDM} on the fact that most experimental attempts assume that axions compose 100\% of the dark matter halo, and an interesting approach is suggested, based on the correlation of two independent experimental results, to look into axion physics without relying on such an assumption.

%%%%%%%%%%%%%%%%%%%%%%%%%%%%%%%%%%%%%%%%%%%%%%%%%%%%%%%%%%%%%%%%%%%%%%%%%%%%%%%%%%%%%%%%%%%%%%
\section{Search Strategies}
\label{sec:strategies}
\begin{figure}
    \centering
    \includegraphics[width=0.9\linewidth]{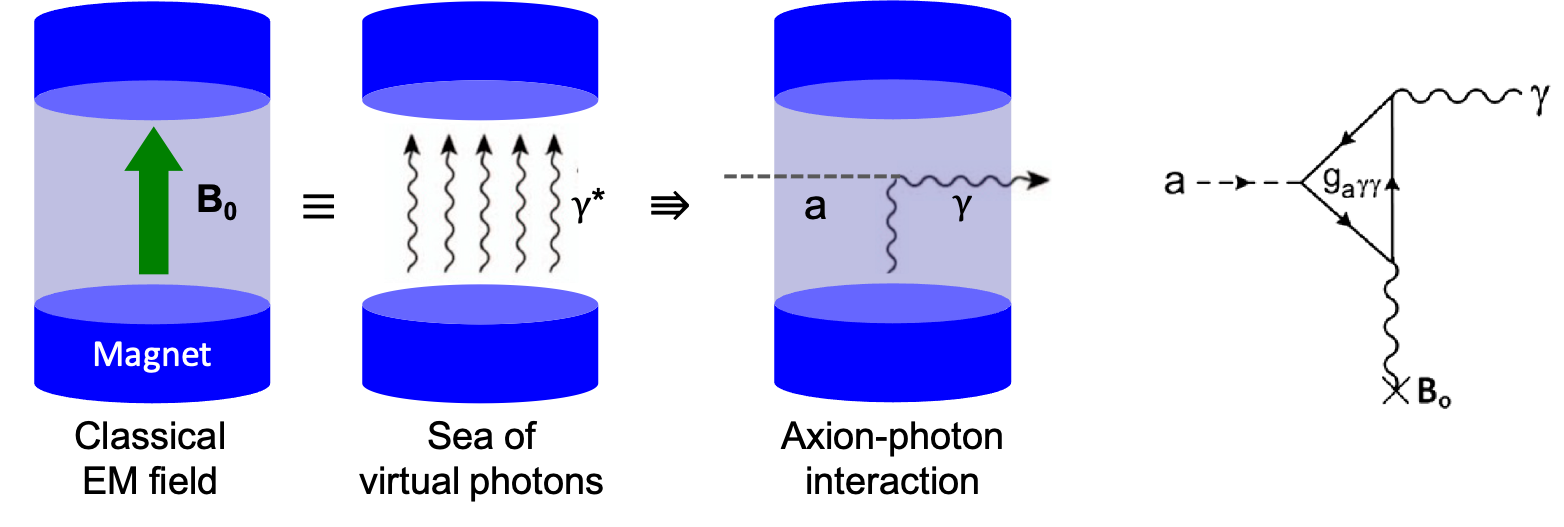}
    \caption{Principle of axion detection. A classical strong magnetic field generates a sea of virtual photons with which axions interact to be converted to real photons. The corresponding Feynman diagram is also shown.}
    \label{fig:primakoff}
\end{figure}

The most common experimental searches for axions rely on the electromagnetic (EM) interaction mediating the axion-photon coupling.
The presence of a coherently oscillating axion field $a$ in free space modifies Gauss' law and Amp$\grave{\rm e}$re's law of Maxwell's equations as
\begin{align*}
    \nabla \cdot \textbf{E} &= \rho -g_{a\gamma\gamma}\nabla a \cdot \textbf{B} \\
    \nabla \times \textbf{B} - \dot{\textbf{E}} &= j + g_{a\gamma\gamma}\left(\dot{a}\textbf{B} + \nabla a \times \textbf{E} \right),
\end{align*}
where $\rho$ and $j$ are the ordinary charge and current while the additional interaction terms, $g_{a\gamma\gamma}\nabla a\cdot\textbf{B}$ and $g_{a\gamma\gamma}(\dot{a}\textbf{B} + \nabla a\times\textbf{E})$, correspond to the axion induced charge and current densities, respectively.
The spatial coherence length (i.e., de Broglie wavelength) of the oscillating axion field is large, e.g. about 1\,km for 1\,$\mu$eV, and inversely proportional to the axion mass.
For invisible axion dark matter, therefore, $\nabla a \approx 0$ is valid and thus only the time dependent current source term is in effect.
In 1983, a promising detection principle was proposed based on the axion-photon interaction to utilize this remaining source term, i.e., the axions are converted into photons in the presence of a magnetic field~\cite{article:Sikivie83}, as illustrated in Fig.~\ref{fig:primakoff}.

First of all, the axions in the $\mu$eV mass range are a plausible cold dark matter candidate and the converted photons are in the microwave frequency region.
The haloscope, a technique for detecting the dark matter halo signal, has provided the most sensitive searches for axions in our galactic halo using microwave resonators.
ADMX, HAYSTAC, and CAPP employ cylindrical cavities in a superconducting solenoid while MADMAX utilizes a periodic array of dielectric planes in a dipole magnet.

Secondly, the axions can emerge abundantly from the stars by Primakoff scattering in the EM fields of charged particles in the stellar plasma.
The helioscope, a dipole magnet directed towards the Sun, is designed to observe X-rays converted from the axion flux originating from the solar core.
The CAST project at CERN has finalized its mission to search for the solar axion in 2015. It will be taken over by a future international program, IAXO, to continue this successful program.

Thirdly, the ALPs can also be sought in the laboratory by converting photons to axions and regenerating them behind a wall based on the so-called light-shining-through-wall (LSW) method.
A combination of high intensity laser beams and strong magnetic fields have been successfully utilized by OSQAR at CERN and ALPS at DESY.
The latter experiment is currently being scaled up with improved sensitivity.

Apart from these, there has been steady progress in the direction of indirect searches for dark matter axions, which are not discussed in this review, using terrestrial radio telescopes to look for spectral lines originating from the resonant conversion of axions to photons by magnetized astrophysical sources, particularly neutron stars~\cite{article:P&P, article:Huang, article:Hook, article:Safdi, article:Leroy, article:Battye}. 
Such radio telescope searches were reported to set constraints on the existence of axion dark matter in the $\mu$eV mass ranges~\cite{article:PSR1, article:PSR2, article:Foster}. 
Recently, major theoretical uncertainties have also been addressed based on advanced simulation and modeling to estimate the expected signal in a more reliable fashion~\cite{article:Chen, article:Carrasco, article:Witte}.

\begin{figure}[t!]
    \centering
    \includegraphics[width=\linewidth]{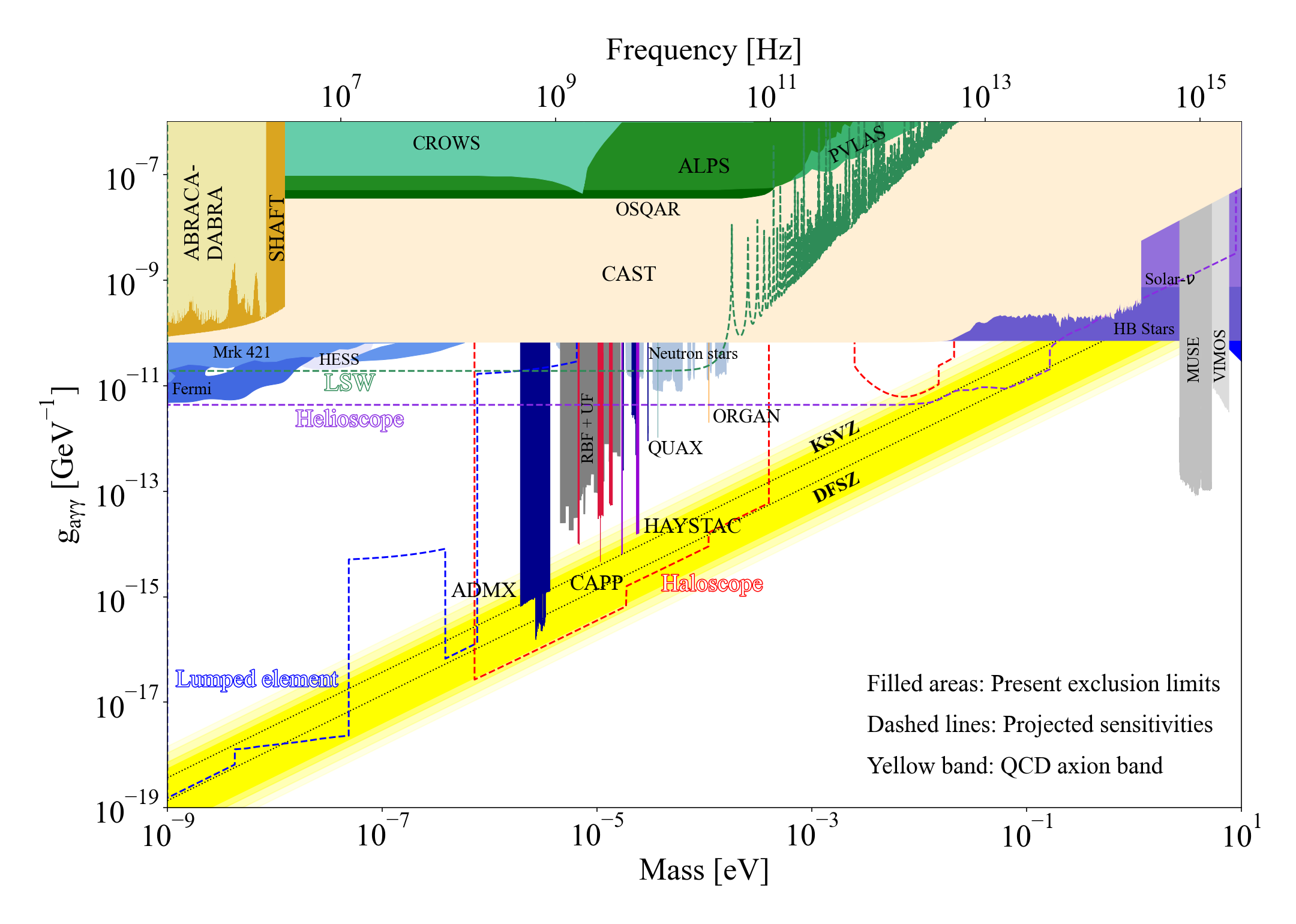}
    \caption{Up-to-date experimental exclusion limits on the axion-photon coupling as a function of axion mass. The projected sensitivities are represented by dashed lines. Two theoretical models are represented by the diagonal dashed lines with the uncertainty band in yellow. Major features of the individual experiments are described in the text.}
    \label{fig:exclusion}
\end{figure}

The search results from these various types of experiments are reflected in a single two-dimensional parameter space defined by the axion coupling to photon, $g_{a\gamma\gamma} = \alpha_{em}C_{\gamma} /{2 \pi f_a}$ (Eq.~16 in the theory section) or equivalently $C_{\gamma}$, and the axion mass $m_a$ or its Compton frequency $\nu_a$ satisfying $m_a \simeq h\nu_a$.
For invisible axions, a linear relation holds between the mass and the Compton frequency, e.g., 1\,$\mu$eV corresponds approximately to 0.25\,GHz.
Experiments utilize diverse detection strategies to explore different mass regions with different sensitivities, which are typically compared with the theoretically interesting axion models, e.g., KSVZ and DFSZ.
Null results exclude the search regions from the parameter space.
Figure~\ref{fig:exclusion} summarizes the exclusion limits on the axion-photon coupling set by individual experiments to date and projected sensitivities for major search methods.
A recent review also discussed search methods for invisible axions and ALPs along with theoretical derivations of their experimental signatures~\cite{article:Sikivie21}, and a committee report provided a comprehensive overview of the experimental programs of direct searches for dark matter candidates~\cite{article:APPEC}.

%%%%%%%%%%%%%%%%%%%%%%%%%%%%%%%%%%%%%%%%%%%%%%%%%%%%%%%%%%%%%%%%%%%%%%%%%%%%%%%%%%%%%%%%%%%%%%
\section{Haloscope Searches}
\label{sec:haloscope}

The axion haloscope was designed to scope microwave photon signals from the axions in our galactic halo, hence the name.
The signal power can be enhanced by the resonance effect that occurs when the axion mass matches the natural frequency of the detection system.
This technique has provided the most sensitive approach to test the theoretical models in the microwave regime.

\subsection{Cavity Haloscopes}\label{subsec:cavity_haloscope}
Microwave cavities possess well-defined eigenmodes (resonant modes) determined by the cavity geometry.
The cavity haloscope employs a microwave cavity resonator immersed in a strong magnetic field.
When the frequency of the axion-induced photon matches the frequency of the cavity eigenmode under consideration, the conversion power is resonantly enhanced by orders of magnitude given by the cavity quality factor $Q_{c}$.
The power deposited in the cavity due to the axion-photon conversion is given, using typical parameter values, by
\begin{equation}
    P_{a\gamma\gamma} = 5.0 \times 10^{-23}\,{\rm W} \left(\frac{C_{\gamma}}{0.75}\right)^2 
    \left(\frac{\rho_a}{0.45\,\frac{\rm GeV}{\rm cm^3}}\right) 
    \left(\frac{\nu_a}{1\,{\rm GHz}}\right) 
    \left(\frac{B_0}{10\,{\rm T}}\right)^2 
    \left(\frac{V}{30\,{\rm L}}\right) 
    \left(\frac{G}{0.5}\right) 
    \left(\frac{Q_c}{10^5}\right),
    \label{eq:conv_power}
\end{equation}
where $C_{\gamma}$ is the model-dependent coupling coefficient with a value of $-$1.92 and 0.75 for the KSVZ and DFSZ model respectively, $\rho_{a}$ is the mass density of dark matter axions as discussed in the theory review, $\nu_a$ is the axion Compton frequency, and $B_0$ is the externally applied magnetic field.
The geometrical factor $G$ is a measure of the amplitude of axion-induced photon field coupled to the cavity mode.

\begin{figure}[t!]
    \centering
    \includegraphics[width=\linewidth]{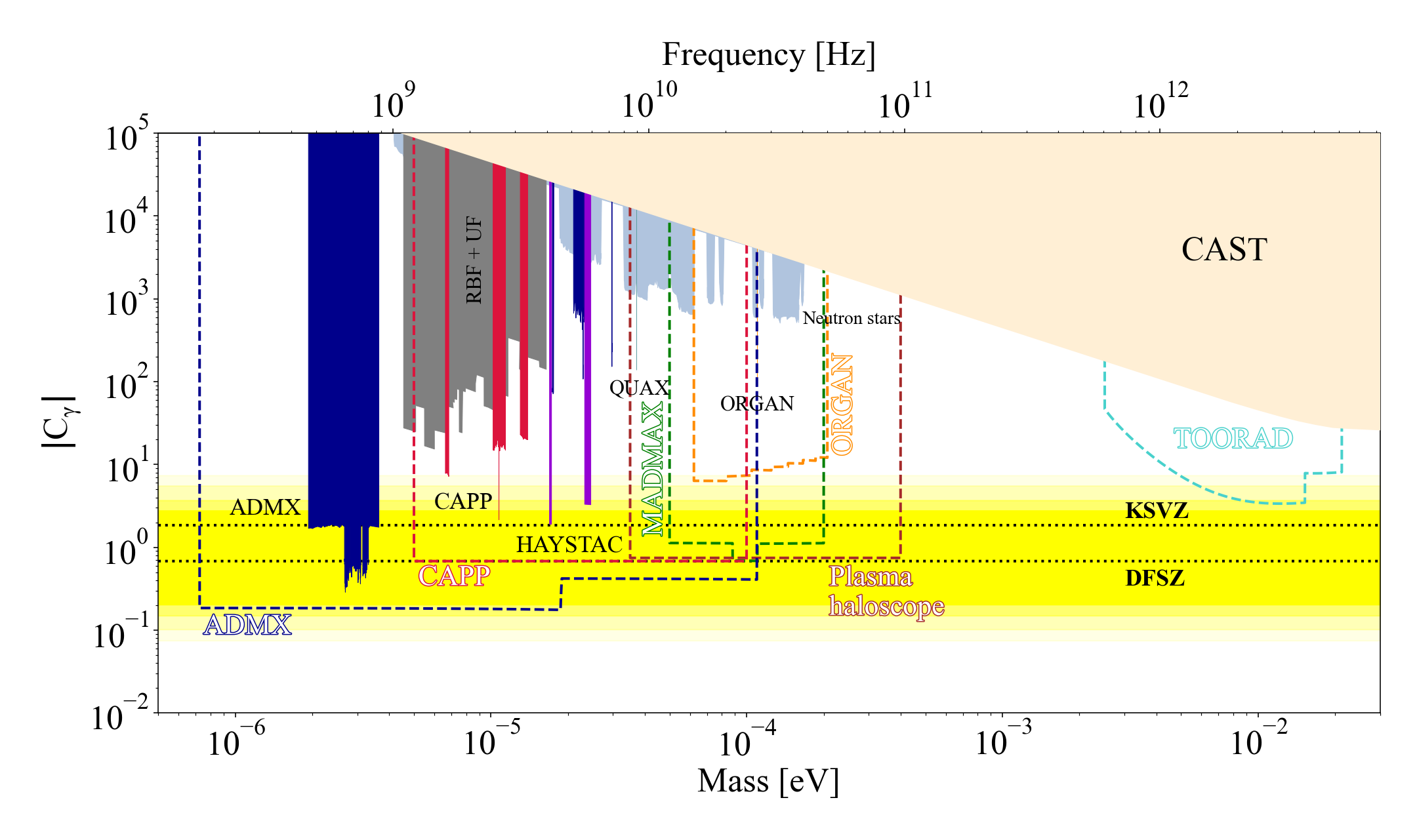}
    \caption{Experimental exclusion limits and projected sensitivities by the haloscopes described in the text in terms of the axion-photon coupling $C_{\gamma}$ as a function of axion mass, presuming the local dark matter halo is made up solely of axions. The projection by CAPP includes the scenario of axions making a partial contribution to the DMHD as discussed in the text.}
    \label{fig:projection_haloscope}
\end{figure}

The experimental sensitivity is determined by the signal-to-noise ratio (SNR),
\begin{equation}
    {\rm SNR} \equiv \frac{P_{a\gamma\gamma}}{\delta P_{\rm sys}},
    \label{eq:snr}
\end{equation}
where $\delta P_{\rm sys}$ is described as fluctuations in system noise power, which is dictated by the Johnson-Nyquist formula, $P_{\rm sys}=k_B T_{\rm sys}\Delta\nu$, with the Boltzmann constant $k_B$ and the equivalent noise temperature $T_{\rm sys}$.
Since the axion mass is a priori unknown, all possible mass ranges need to be explored.
A relevant figure of merit for experimental design is the scanning rate, i.e., how fast one can scan a mass region with a given sensitivity.
This quantity is obtained by plugging Eq.~\ref{eq:conv_power} into Eq.~\ref{eq:snr} as
\begin{equation}
    \frac{d\nu}{dt} = 1.2\,{\rm \frac{GHz}{year}} 
    \left(\frac{5}{snr}\right)^2 \left(\frac{0.15\,{\rm K}}{T_{\rm sys}}\right)^2
    \left(\frac{P_{a\gamma\gamma}(B_0,V,G,Q_c)}{5.0\times10^{-23}\,{\rm W}}\right)^2
    \left(\frac{10^5}{Q_c}\right),
\end{equation}
where $snr$ is the target SNR value.
Major R\&D efforts are made to increase the scanning rate by maximizing $B_0^2VGQ_c$ and minimizing $T_{\rm sys}$. 
The up-to-date exclusion limits made by individual haloscope experiments, which will be described in detail in this section, and their projected sensitivities are shown in Fig.~\ref{fig:projection_haloscope}.

\begin{comment}
The most sensitive experimental approaches to date are based on the ``haloscope" method~\cite{article:Sikivie83} utilizing a strong  magnetic field, a high quality microwave resonator, and a low-noise cryogenic receiver. 
However, the advanced searches (see Sec.~\ref{subsec:cavity_haloscope}) have covered a small part of the theoretically interesting regions, with most of the available parameter space unexplored. 
\end{comment}

The first experimental search for axion dark matter was carried out by the Rochester-Brookhaven-Fermilab (RBF) Collaboration using a SC solenoid magnet, a copper cavity in a liquid helium reservoir and a transistor-based RF-amplifier.
The sensitivity, however, was not high enough to reach the axion theory band.
The system noise was reduced by a subsequent experiment at the University of Florida (UF) to make the search more effective.
The Axion Dark Matter eXperiment (ADMX), established in 1990, designed a larger scale detector to explore new ways forward.
Several critical improvements were involved to increase the sensitivity to the theoretically interesting levels.
These include the employment of a SC quantum interference device (SQUID) and/or a Josephson parametric amplifier (JPA) to reduce the electronic noise near to the quantum limit; and the operation of a dilution refrigerator to lower the system temperature to the sub-Kelvin level.
However, the advantages of such a large-scale detector with the quantum-noise-limited technology would run out quickly at higher frequencies because the large cavity diameter limits the search to low frequencies below 1\,GHz and also the quantum noise grows linearly with frequency.
In addition, overall progress in the axion field has been time-consuming and one would think that the task is just too daunting to carry out. 
The IBS Center for Axion and Precision Physics Research (IBS-CAPP) launched a massive, parallel R\&D program to push the limits of current technology in order to extend the search range towards higher frequency regions with enhanced detection efficiency.
The breakthroughs contain 1) the successful operation of multiple experiments simultaneously; 2) the development of functional high temperature superconducting (HTS) and low temperature superconducting (LTS) magnets capable of reaching higher than 9T (the limiting value of traditional LTS-magnets); 3) the development of efficient high-frequency microwave resonators with large volume; 4) developed practical superconducting cavities based on HTS tapes with large quality factors; and 5) the full use of the most effective quantum noise limited amplifiers.
In the mean time, the cavity haloscope searches based on cylindrical cavities in a SC solenoid became popular such that many other collaborations have proposed and demonstrated their new ideas.
Separate efforts, on the other hand, have been made to recycle the existing resources, such as the prototype SC dipole magnets at CERN, for axion dark matter search by employing arrays of small rectangular cavities placed inside the long magnet bores.

\subsubsection{RBF-UF}
The first microwave cavity experiment looking for axion dark matter was conducted by the RBF Collaboration at Brookhaven National Laboratory in 1987~\cite{article:RBF1}.
The experimental setup, providing the basic structure for even today's most sensitive experiments, includes a SC solenoid magnet with a maximum field of 8.5\,T, a copper resonant cavity immersed into liquid helium, heterostructure field-effect transistor (HFET) RF-amplifiers, and a computer-aided automated data acquisition system.
The cavity was operated at two different resonant modes tuned by insertion of a dielectric or conducting rod, accompanying a study of mode localization.
Several experimental configurations set limits in the range $4.5\,\mu{\rm eV} < m_a < 16.3\,\mu$eV on the axion-photon coupling.
The sensitivity level was about a factor of $10^2$ to $10^3$ away from the QCD axion models mainly due to the high system noise temperature of about 16\,K~\cite{article:RBF2}.
A subsequent search performed at the University of Florida (UF) was rapidly reported with reduced system noise using a commercially available amplifier with a noise temperature of about 3\,K, which improved sensitivity by an order of magnitude for masses around 5.5\,$\mu$eV~\cite{article:UF}.

\subsubsection{ADMX}
ADMX is the first large-scale cavity experiment searching for QCD dark matter axions with realistic masses and couplings to microwave photons.
The detector consists of a cylindrical tunable high-$Q$ microwave cavity placed in a 7.6\,T magnetic field produced by a 0.6\,m wide and 1\,m long SC solenoid.
The large size of the apparatus featuring a high value of $B^2V$ naturally fits sensitive searches for axions in relatively low mass regions, particularly between 1\,$\mu$eV and 10\,$\mu$eV.
During the first operation period, the detector was cooled down to 1.5\,K and the electromagnetic power spectrum was obtained by a low noise microwave receiver.
The experiment has achieved a power sensitivity better than 10$^{-23}$\,W to probe the KSVZ axions in the mass range 1.9$-$3.3\,$\mu$eV~\cite{article:ADMX_KSVZ1, article:ADMX_KSVZ2}.
The subsequent run implemented a quantum technology, SQUID, whose intrinsic noise is subject to the standard quantum limit.
A frequency tunable microstrip SQUID amplifier replacing the HFET amplifier has broadened the search range with an improved detection rate by 2 orders of magnitude~\cite{article:ADMX_SQUID}.

The second generation detector is featured by deploying  a dilution refrigerator, enabling the experiment to operate at sub-Kelvin temperatures.
A substantial reduction in thermal background noise as well as the excess noise from the quantum-limited SQUID amplifier resulted in many-fold improvement in the system noise temperature providing the capability to achieve unprecedented sensitivity to the DFSZ model for the first time~\cite{article:ADMX_Run1A}.
Recently, they extended the search range to higher frequencies by utilizing an ultralow-noise JPA with a tunable resonance developed by the University of California, Berkeley~\cite{article:ADMX_Run1B}.
Ongoing searches are expected to provide crucial tests of the invisible axion models over a wider mass range.

Meanwhile, ADMX has been putting significant research and development effort to expand searches beyond 10\,$\mu$eV.
A prototype of a small-scale cavity mounted on top of the main cavity system exploited multimode searches to attain new mass limits in three distinct frequency regions~\cite{article:ADMX_sidecar}.
A multiple-cavity system is currently under construction to compensate for volume reduction in high-mass axion searches with a refined design from the proof of concept study performed earlier using an array of four cavities~\cite{thesis:multicavity}.
The principal challenge of this design, maintaining frequency-matching among the cavities in every frequency step, needs to be carefully addressed and realistically demonstrated~\cite{article:multicavity}.

To further increase sensitivity to the microwave signal detection, the Fermilab group has been developing a qubit-based photon counting technique which allows for repeated non-demolition measurements of cavity photons to reduce the noise~\cite{article:ADMX_SPD}.
They claimed a noise reduction of more than 15\,dB at near 25\,$\mu$eV (6.0\,GHz) below the quantum limit with the detector performance being limited by residual background of real photons.
A new experimental exclusion limit on hidden photon dark matter has proven that this novel technique can greatly enhance the searching speed for dark matter axions.

\subsubsection{HAYSTAC}
The HAYSTAC (Haloscope At Yale Sensitive To Axion CDM) experiment was designed to search for dark matter axions with masses above $20\,\mu$eV, an order of magnitude higher in mass than naturally accessible by ADMX.
A 2L copper-plated stainless steel cylindrical cavity is housed in a cryogen-free dilution refrigerator integrated with a 9T SC solenoid.
HAYSTAC is the first microwave cavity haloscope search with deployment of a JPA, developed by JILA (the Joint Institute for Laboratory Astrophysics at the University of Colorado), that owes its inductance to an array of SQUIDs to achieve near-quantum limited performance.
The parametric amplification is attained by a coherent pump field that modulates the resonant frequency, which can be tuned by DC flux through the SQUID array.
The experiment has completed its first data production at around 24\,$\mu$eV and excluded the possibility of axion conversion into two photons at slightly higher level than the KSVZ coupling~\cite{article:HAYSTAC1,article:HAYSTAC2}, demonstrating that a small-scale axion haloscope will still be able to probe the axion physics at higher mass regions.

Recently, the Collaboration used a pair of JPAs to realize squeezed vacuum states to circumvent the fundamental quantum-noise limit.
One of them was dedicated to prepare a photon field in a squeezed state, while the other was used to read out only the squeezed quadrature with nearly noiseless amplification.
This quantum squeezing technique effectively increased the bandwidth over which the apparatus is sensitive to an axion obtained at a given cavity resonant frequency.
By coupling the HAYSTAC cavity to the squeezed-state receiver, they yielded a twofold enhancement in scanning rate relative to optimal unsqueezed operation and set a new limit on axion coupling at around 4\,GHz (17\,$\mu$eV)~\cite{article:squeezing}.

\subsubsection{CAPP}\label{sec:capp}

IBS-CAPP was established with the aim of building a laboratory equipped with the world’s best infrastructure for cavity haloscope searches with enhanced sensitivities over a boarder range in the microwave region. 
The laboratory is featured by multiple state-of-the-art axion experiments $-$ three already built and in operation mode, one being under construction, and another in the planning stage $-$ based on high field SC magnets, powerful dilution refrigerators mounted on low vibration pads, and quantum noise limited amplifiers.
The individual experiments were designed to run in parallel targeting at different mass ranges in order to explore the axion parameter space as quickly as possible.

Since the field of axion dark matter search requires highly-developed technologies as well as advanced scientific skills from a number of different research areas, the CAPP's strategies were set accordingly early on with a major emphasis placed in the Center's devotion to diverse R\&D areas rather than a single project.
The ongoing R\&D endeavors focus on:
\begin{itemize}
    \item Launch of a pioneering program of axion dark matter search using HTS magnets based on ReBCO tapes, capable of withstanding fields up to 100\,T with adequate current density~\cite{article:25T_magnet}.
    \item Establishment of an experiment for definitive tests of the QCD axion models above 1\,GHz making use of the high field and large volume of a 12T/320mm low temperature SC (LTS) magnet and ultra-low temperature achievable by a powerful dilution refrigerator.
    \item Design and construction of high-efficiency high-frequency microwave resonators in the most advantageous ways~\cite{article:pizza_cavity}. 
    \item Development of practical superconducting resonators (SCRs) with quality factors larger than normal conductor by at least an order of magnitude~\cite{article:SC_cavity}.
    \item Optimal use of a series of the most effective JPAs (and traveling wave JPAs (TWJPAs) if available) covering from 1\,GHz to 8\,GHz to bring the electronic noise  near the quantum limit~\cite{article:CAPP_JPA}.
    \item Implementation of a phase-matching (PM) scheme $-$ coherent combination of individual signals from multiple local systems running at the same frequency.
    \item Finally, development of a tunable microwave-photon detector based on quantum technologies using Rydberg atoms particularly for high frequency searches above 10\,GHz.
\end{itemize}

At the current stage, the ground at CAPP was prepared for sensitive searches for axions in the 5$-$30\,$\mu$eV mass range on a timescale of five years, presuming dark matter in the local halo is solely made up of axions. 
Recently, three experiments, CAPP-8TB, CAPP-9T, and CAPP-PACE, have carried out their pioneering operation to yield meaningful scientific results around 6.7\,$\mu$eV, 13.5\,$\mu$eV, and 10.7\,$\mu$eV, respectively~\cite{article:CAPP-8TB, article:CAPP-MC, article:CAPP-PACE}.
CAPP-8TB and CAPP-PACE set up similar experimental designs consisting of a 8T SC magnet installed in a dilution refrigerator to produce the first result and scan more than 1\,$\mu$eV with sensitivities close to the KSVZ model.
CAPP-9T used a 9T SC magnet and a He-3 refrigerator to demonstrate the performance of the new cavity concept, featured by a multiple-cell structure in a single cavity~\cite{article:pizza_cavity}, which will promote high-mass axion searches.
These experiments are currently being upgraded for improved sensitivities with adoption of high-performance JPAs from the University of Tokyo and implementation of high-Q cavities uniquely developed by CAPP using HTS tapes and/or dielectric arrays.

The CAPP-12TB experiment, presently under commissioning, will take advantage of the high field and large volume of the customized LTS-12T/320mm magnet, which was fully tested in 2020, expecting to lead the cavity haloscope search program.
The large aperture of the magnet allows for nearly 40 liters of detection volume and a powerful dilution refrigerator brings the experimental system to an extreme environment below 100\,mK. 
Using a series of JPAs with different dynamic ranges, CAPP-12TB is expected to deliver unprecedented experimental sensitivities above 1~GHz, probing the DFSZ axion physics, in a wider mass range.
An additional experiment plans to join the parallel search program in 2022 to extend the search range beyond 10\,GHz utilizing a 12T SC magnet with a smaller aperture. 
With successful integration of the R\&D efforts into these experimental setups, CAPP desires in the next decade to be sensitive enough to examine the supposition that the QCD axion partially composes the local dark matter halo density, over a range between 1\,GHz and 25\,GHz (4\,$\mu$eV and 100\,$\mu$eV).

\subsubsection{Other cavity haloscopes}
The QUAX (QUest for AXions) Collaboration puts an effort to bring a cavity experiment mining the axion-photon coupling around 10\,GHz. 
In particular, they performed an axion search at 9\,GHz employing a NbTi SC resonant cavity to achieve a quality factor $Q_c=4\times10^5$, a factor 4 higher than a copper cavity, under a 2T magnetic field.
This corresponds to the first search for galactic axions using a SC haloscope~\cite{article:QUAX_SCC}.
The Collaboration also simulated a single photon counter based on an underdamped Josephson junction, claiming that the switching voltage measurement can register microwave single photons with a dark count rate of $4\times10^{-4}\,$Hz~\cite{article:QUAX_SPD}.

An Australian group designs an experiment, named ORGAN (Oscillating Resonant Group AxioN), to probe high-mass regions considering a unique cavity configuration~\cite{article:ORGAN}.
Consisting of a variety of thin, long cylindrical cavities of various dimensions packed in a 14T SC magnet, similar in arrangement to a pipe organ, the experiment  desires to access different regions in a single operation to cover a wide range of 60$-$200\,$\mu$eV 

A Large Hadron Collider test dipole magnet, which had been used to look for solar axions at CERN, was recycled for dark matter axion searches benefiting from the high field and large length of the magnet.
A joint project, CAST-CAPP, placed a series of 40cm-long tunable rectangular cavities inside one of the magnet twin bores to exploit the phase-matching scheme for an initial dark matter axion search above 5\,GHz~\cite{proc:CAST-CAPP}.
The RADES (Relic Axion Detector Exploratory Setup) project plans to utilize the same dipole magnet to explore masses of several tens of $\mu$eV using an array of small microwave cavities connected by rectangular irises, whose characteristics were recently studied~\cite{article:RADES}.

\subsection{Dielectric Haloscopes}
Relying on closed resonant structures, microwave cavity haloscopes are sensitive to dark matter axions with relatively low frequencies up to several GHz, but they would face difficulties in scaling to higher frequencies for which the cavity size needs to be smaller.
An alternative method for efficient searches for high-frequency axions is based on a detector architecture consisting of strategically configured dielectric materials in an open resonator.
In particular, periodic structures of dielectric planes with a high dielectric constant $\epsilon_r$ are expected to allow for an observable emission of electromagnetic waves induced by axions with frequencies (masses) between 10 (40) and 100 (400)\,GHz ($\mu$eV).

\subsubsection{Open resonator}
The first version of an open resonant system introduced a series of current-carrying wire planes in an RF Fabry-Perot resonator which provides an alternating magnetic field with the same polarity as the axion-induced electric field to maximize the geometry factor.
The Orpheus experiment demonstrated this technique using copper wire planes for a search for dark matter ALPs with masses around 70\,$\mu$eV~\cite{article:Orpheus}.
A new design was proposed with the wire planes replaced by high-$\epsilon_r$ dielectric planes to modify the axion-induced electric field yielding non-vanishing geometry factors under a static magnetic field.
These two approaches are conceptually similar to each other, but the latter is more beneficial in acquiring higher quality factors.

\subsubsection{MADMAX}
The MADMAX (Magnetized Disc and Mirror Axion eXperiment) Collaboration proposed a new strategy to search for dark matter axions which can explore a similar mass range $40-400\,\mu$eV~\cite{article:MADMAX, article:Brun19}. 
The scheme is characterized by the so-called ``booster" which is composed of a periodic structure of high-$\epsilon_r$ dielectric disks with a RF mirror on one side, all of which are placed in an external magnetic field parallel to the surfaces.
The changing dielectric media induce discontinuity in the axion-induced electric field in the interface, resulting in generation of EM waves propagating perpendicular to the dielectric surface in both directions.
By exploiting constructive interference of the emitted waves from the precisely aligned interfaces and the reflected waves from the mirror, the radiated electromagnetic signal is boosted, hence the name, by the number of disks in quadrature.
Analogous to the quality factor in a cavity haloscope, this boosting effect acts as an enhancement factor in a dielectric haloscope, which is a nonresonant open system.
The expected signal power from such a boosting haloscope received by an antenna placed on the other side of the system is given by
\begin{equation}
    P=1.3\times10^{-21}\,W 
    \left(\frac{B_0}{10\,{\rm T}}\right)^2 
    \left(\frac{C_{\gamma}}{1.92}\right)^2
    \left(\frac{\beta}{400}\right)^2 
    \left(\frac{A}{1\,{\rm m^2}}\right),
    \label{eq:power_madmax}
\end{equation}
where $\beta$ is the boosting factor and $A$ is the disk surface area.
Since the search frequency is determined by the space between neighboring disks, the detector configuration facilitates access to high frequency regions without major loss of detection volume in contrast to conventional cavity experiments.

For realistic sensitivity to reach the QCD axion models, the Collaboration proposed to build a large scale dielectric haloscope experiment which is featured by a 2m-long 10T dipole magnet, a boosting system consisting of 80 dielectric disks with 1m$^2$ area, and a low noise receiver chain.
Recently, they demonstrated a first proof of principle realization of the booster setup of the haloscope using a copper mirror and up to five sapphire disks~\cite{article:booster} and performed a dedicated 3D simulation study to investigate design requirements for the dielectric disk system~\cite{article:MADMAX-3D}. 

\subsection{Other Haloscopes}

\subsubsection{Ferromagnetic haloscope}
Another experiment in the QUAX Collaboration accepts a variant of cavity haloscope, named QUAX-$ae$, exploiting the interaction of the cosmic axion with the spin of electrons in a magnetized sample placed inside a resonant cavity.
The resonant interaction of the axion-induced EM field at the Larmor frequency, determined by an external magnetic field, would flip the electron spin states (magnons) coherently which subsequently emits microwave photons~\cite{article:QUAX_proposal}. 
QUAX-$ae$ is unique in probing the axion-electron coupling directly, giving excellent prospects for model discrimination in the event of discovery.
A preliminary axion dark matter search demonstrated the scheme using a photon-magnon hybrid system consisting of a series of Yttrium iron garnet spheres coupled to the TM$_{110}$ mode of a cylindrical copper cavity~\cite{article:QUAX-ae_1}.
A subsequent scientific run with an upgraded photon-magnon system coupled to a quantum-limited Josephson parametric amplifier resulted in the best limit on the axion-electron coupling near 43\,$\mu$eV~\cite{article:QUAX-ae_2}.
The Collaboration also plans further upgrades using 100's of spheres with the goal of DFSZ sensitivity.

\subsubsection{Plasma haloscope}
A proposed search strategy for dark matter axions, known as plasma haloscope, considers the coupling of the axions to bulk plasmons, quanta of plasma oscillation, rather than photons~\cite{article:plasma}.
It utilizes the resonant conversion between the two particles by matching the axion mass to the plasma frequency.
A metastructure of conducting wires is identified as a promising candidate for tunable plasma with the frequency varying with interwire space. 
A key advantage of this concept is that the resonant frequency and hence the signal enhancement is unrelated to boundary conditions, allowing for a large conversion volume over a wide frequency range.
For realistic experimental sizes, plasma haloscopes can offer a plausible alternative to dielectric haloscopes providing competitive sensitivity in the mass region $35-400\,\mu$eV.

\subsubsection{Topological Insulator}
An experimental idea was suggested based on topological phenomena in condensed matter physics.
The existence of hypothetical axion-like quasiparticles in topological insulators allows the mathematical description to be identical to axion electrodynamics.
Antiferromagnetically doped topological insulators predict such dynamical quasiparticles that are resonantly driven by the axion-induced field under an external magnetic background and subsequently converted into photons in the THz range~\cite{article:AQ-TI}.
This presents a viable route to detect axion dark matter with mass of order 1\,meV, currently inaccessible by other dark matter detection methods.
The independence of sample volume of the resonance allows for high sensitivity at high frequencies with broad tunability provided by varying the external magnetic field. 
A recent proposal called TOORAD (TOpolOgical Reseonant Axion Detection) has developed the theory of axion quasiparticles in topological magnetic insulators and characterized a realistic experimental setup with an antiferromagnetic insulator (Fe-doped Bi$_2$Se$_3$ or Mn$_2$Bi$_2$Te$_5$) to realize such axion quasiparticles assuming a wide bandwidth single photon detector highly efficient in THz~\cite{article:TOORAD}.

\subsection{Low Mass Axion Searches}
Complementary to the cavity or dielectric haloscope method, searches for axions in low mass regions $m_a < 10^{-6}$\,eV require different strategies.
Dark matter axions can cause an oscillating electric current by an external magnetic field which subsequently induces a small magnetic field.
This axion-induced field can be detected exploiting a lumped element circuit and a sensitive magnetometer in order to probe axion-photon couplings weaker than those accessible by haloscope searches~\cite{article:LC_circuit}.
Magnetic resonance techniques can also be used to detect a time-varying torque on nuclear spins exerted by an oscillating axion field and be sensitive to ALP dark matter with masses $m_a < 10^{-9}$\,eV~\cite{article:CASPEr}.

\subsubsection{Lumped element searches}
An $LC$ circuit based axion search was performed by the pilot experiment ADMX SLIC (Superconducting $LC$ Circuit Investigating Cold Axions), which was designed to probe lighter-axion mass parameter space difficult to reach with microwave cavity haloscopes.
The prototype circuit consists of a large rectangular SC loop antenna placed in a solenoid magnet bore and a parallel plate capacitor with a movable dielectric sheet in between to tune the resonant frequency.
The experiment demonstrated this new strategy by exploring three consecutive mass regions near $10^{-7}$\,eV with different magnetic fields.

ABRACADABRA (A Broadband/Resonant Approach to Cosmic Axion Detection with an Amplifying $B$-field Ring Apparatus) is an experiment proposed to search for axion-photon coupling over a broad mass range $10^{-12}\,{\rm eV} < m_a < 10^{-6}$\,eV~\cite{article:ABRACADABRA}. 
It employs a static toroidal magnetic background to source a time-varying effective electric current along the magnetic field, $\textbf{J}_{\rm eff} = g_{a\gamma}\sqrt{2\rho_a}\cos{(m_at)}\textbf{B}$, that gives rise to an oscillating magnetic flux through the center of the toroid, which can eventually be sensed by a SQUID magnetometer.
The readout circuit can be designed for resonant/broadband searches with/without a tunable capacity in the input coil.
Recently, a small-scale prototype demonstrated the detection scheme which made an advance towards a full-scale experiment that can examine the QCD axions with low masses~\cite{article:ABRACADABRA-10cm}.

A dual search program, DM Radio (Dark Matter Radio), for axion and hidden photon dark matter is a solenoid version of the lumped element search~\cite{article:DM_Radio}.
Using an optimized tunable SC $LC$ resonator, it found the potential sensitivity of many orders of magnitude beyond current limits over an extensive frequency range from 100\,Hz to 300\,MHz.
The prototype DM Radio pathfinder experiment, desined to probe hidden photons in the 100\,kHz$-$10\,MHz mass range, is currently under construction~\cite{article:DM_Radio_design}.
The full-scale program plans to instrument a sample of $\sim$1\,m$^3$ at 10\,mK to test the pre-inflation dark matter scenario.

SHAFT (Search for Halo Axions with Ferromagnetic Toroids) is a new experimental program utilizing a similar concept to ABRACADABRA.
A main difference is to make use of toroidal magnets with ferromagnetic material in the core in order to enhance the magnetic field. 
The experimental apparatus is configured with two pairs of stacked toroids each of which has a separate pickup coil a SQUID magnetic flux sensor.
The toroid pairs generate magnetic fields in the opposite direction such that two independent detection channels reduce the correlated systematic noise.
The experiment was able to reach a magnetic field of 1.5\,T at 6\,A using a set of toroids with Fe-Ni powder cores and achieved magnetic sensitivity of 150\,aT/$\sqrt{\rm Hz}$ using a dual-channel readout chain at 4.2 K.
Based on this configuration, they reported on a direct search for EM interactions of ALPs in a broad low-mass region that spans from $10^{-11}$ to $10^{-8}$\,eV~\cite{article:SHAFT}.

\subsubsection{NMR-based search}
The presence of the oscillating cosmic axion background could produce a time-varying nuclear electric dipole moment (EDM)~\cite{article:CASPEr}. 
If the axion-induced EDM oscillates at the nuclear Larmor frequency, this will result in the enhancement of the transverse magnetization with time that can be measured with a sensitive magnetometer such as SQUID.
CASPEr (Cosmic Axion Spin Precession Experiment) is a nuclear magnetic resonance experiment seeking for such oscillating EDM signals using two different measurement schemes via different couplings.
CASPEr-Wind, the originally proposed scheme, searches for dark matter ALPs by considering their pseudo-magnetic coupling to nucleons, referred as the ALP-nucleon coupling.
The scheme offers a resonant search via continuous-wave NMR spectroscopy and yields the highest sensitivity for frequencies ranging from a few Hz to hundreds of MHz, corresponding to masses $10^{-14}$ to $10^{-6}$\,eV.
A new search method, called CASPEr-Electric, on the other hand, is sensitive to experimental signatures of the axion-gluon coupling.
The concept makes use of a static electric field to implement a non-resonant frequency-modulation detection scheme, suitable for searches below Hz down to mHz (masses $10^{-17}$ to $10^{-14}$\,eV), extending the detection bandwidth by three decades.

A similar approach is considered with storage ring experiments designed to measure the proton EDM by looking for the time-varying spin precession induced by ultralight dark matter relying on the axion-gluon coupling.
A combination of external magnetic and electric fields tunes the $g-2$ frequency of polarized proton beams to be in resonance with the oscillation frequency of the background axion field, which brings the enhancement in the oscillation amplitude of the axion-induced EDM with time~\cite{article:Chang19}.
The proton is given the so-called magic momentum to keep the spin and momentum always aligned, and if the proton has an EDM, its spin will precess in the presence of an external electric field.
The effect will add up for each proton orbit around the ring and boost the signal over the entire particle storage time in the ring, about 1000\,s for the proton EDM experiment~\cite{article:Graham21}.
Since a relativistic proton sees a much larger spatial gradient of the axion field, the storage ring EDM method can in principle provide a sensitive probe particularly to low-mass regions from $10^{-24}$ to $10^{-7}$\,eV.

\begin{comment}
\begin{figure}
    \centering
    \includegraphics[width=.6\linewidth]{CASPEr_wind.png}
    \caption{Geometry of the CASPEr experiment. The Larmor frequency of the nuclear dipole moment depends on the applied magnetic field. When this frequency matches the axion dark matter frequency, a transverse polarization will be built-up, which can be detected by the SQUID pickup loop.}
    \label{fig:CASPEr_wind}
\end{figure}

\begin{figure}
    \centering
    \includegraphics[width=1.\linewidth]{CASPEr_schedule.png}
    \caption{Estimated experimental reach of CASPEr. It plans to cover up to close to 1\,MHz range already from phase II and eventually reach 10\,MHz with phase III all the way down to QCD axion sensitivity (Adapted from the presentation by Dima Budker, originally provided by Alexander Sushkov, at the IBS Conference on Dark World in Daejeon, November 2019).}
    \label{fig:CASPEr_schedule}
\end{figure}
\end{comment}

%%%%%%%%%%%%%%%%%%%%%%%%%%%%%%%%%%%%%%%%%%%%%%%%%%%%%%%%%%%%%%%%%%%%%%%%%%%%%%%%%%%%%%%%%%%%%%
\section{Helioscope Searches}\label{sec:helioscope}
\label{sec:helioscope}

The high-mass regions have been looked into by either astrophysical telescopes or laboratory based experiments. 
As a matter of fact, most of the regions where the axion could exist without being dark matter have been already excluded, so it is not unnatural to assume that it may constitute the dark matter in our galactic halo. 
This assumption, however plausible as it may be, is still an assumption. 
If the axion physics takes place in Nature, stars are the strongest sources of the axion production via the Primakoff conversion of the plasma photons, reducing the uncertainty about their production mechanism. 
Helioscopes were designed to detect the copious flux of axions emitted from our Sun based on the axion-to-photon conversion using a dipole magnet directed toward the Sun.
No matter what the axion mass is, the total axion energy would reflect the temperature of the Sun’s interior of a few keV, as they would be efficiently produced in the  solar core within 10\% of the solar radius.
Therefore, the solar axions searches utilize photon detectors with high efficiency in the X-ray region.
The probability of axions conversion into photons is given by 
\begin{equation}
    %\mathcal{P}_{a\rightarrow\gamma} = \left(\frac{g_{a\gamma\gamma}B_0}{q}\right)^2 \sin\left(\frac{qL}{2}\right),
    \mathcal{P}_{a\rightarrow\gamma} = 2.6\times10^{-17} \left(\frac{g_{a\gamma\gamma}}{10^{-10}\,{\rm GeV^{-1}}}\right)^2 \left(\frac{B_0}{10\,T}\right)^2 \left(\frac{L}{10\,{\rm m}}\right)^2 \mathcal{F},
\end{equation}
where $L$ is the magnet length.
The enhancement factor $\mathcal{F}$ accounts for the coherence of the process, whose value is preserved ($\mathcal{F}\simeq1$) for axion masses up to $\sim$10$^{-2}$\,eV.

\begin{figure}
    \centering
    \includegraphics[width=\linewidth]{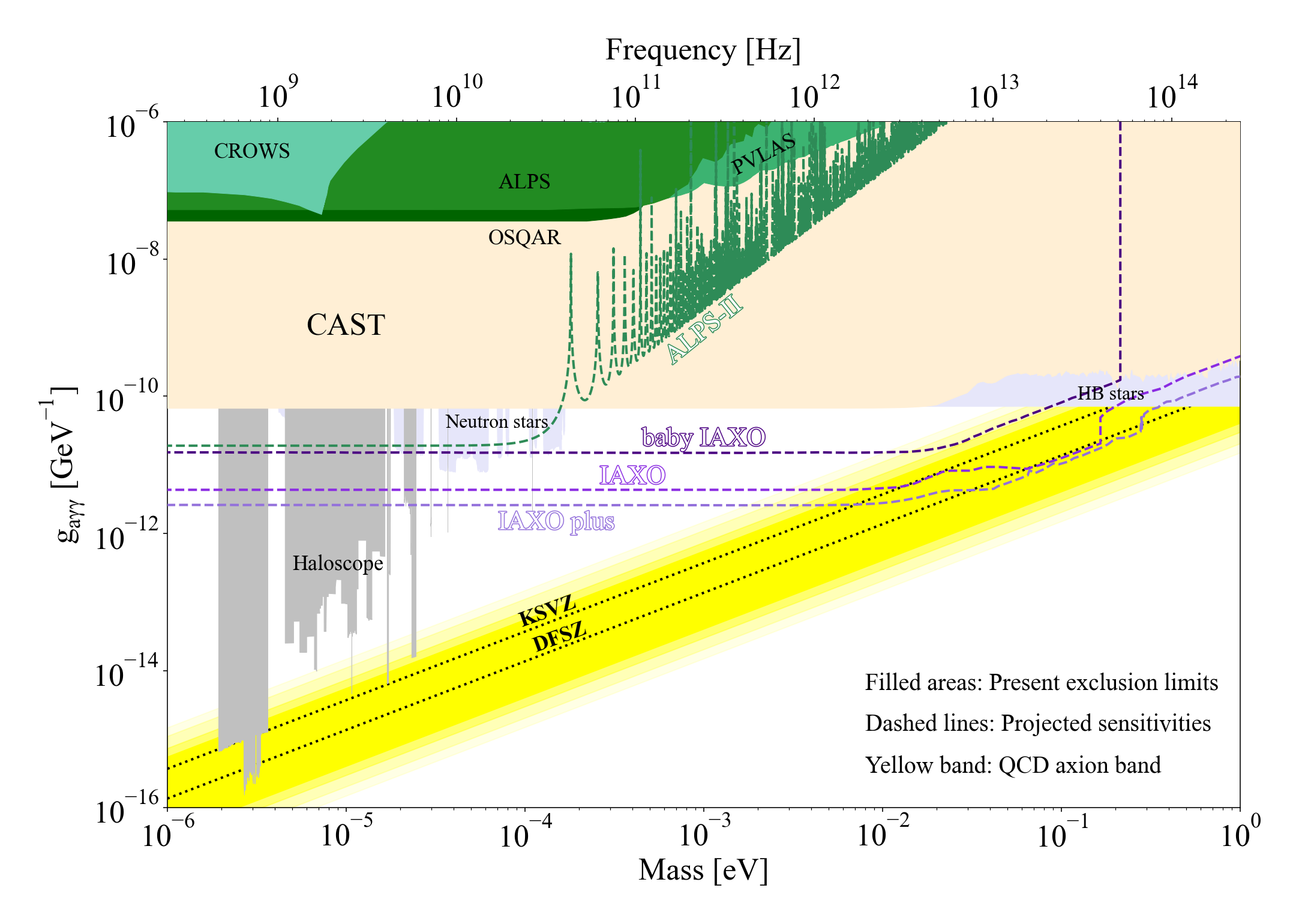}
    \caption{Present exclusion limits by the past helioscope (CAST) and LSW experiments (OSQAR and ALPS).
    The projected sensitivities from their next generation experiments (IAXO and ALPS II) are represented by the dashed purple and green lines.}
    \label{fig:sensitivity_helio-LSW}
\end{figure}

Historically, the first helioscope search was performed at Brookhaven National Laboratory~\cite{article:BNL_Helioscope}, then at the University of Tokyo~\cite{article:Tokyo_Helioscope}, and finally at CERN with CAST (CERN Axion Solar Telescope)~\cite{article:CAST}. 
CAST recycled a LHC dipole prototype magnet with a magnetic field of up to 9\,T over a length of 9.3\,m and has been operational for more than a decade.
With improved detectors and novel X-ray optics, the final phase of the experiment set the most restrictive constraint on the axion-photon coupling below $10^{-10}$\,GeV$^{-1}$ for masses $m_a<0.02$\,eV, competing with the most stringent limits from astrophysics~\cite{article:CAST}.
IAXO (International Axion Observatory), the fourth generation axion helioscope, was proposed to improve CAST's performance by more than an order of magnitude by building a large-scale magnet and X-ray focusing devices coupled to low background detectors~\cite{article:IAXO1, article:IAXO2}.
The new magnet design, inspired by the ATLAS toroids~\cite{article:ATLAS}, considers a multibore configuration in a large toroid comprising eight 21m long and 1m wide racetrack coils, which can generate an intense magnetic field of 5.4\,T inside eight 600\,mm aperture bores placed between two neighboring coils.
Baby-IAXO, a demonstration version of the proposed large-scale IAXO with a 10\,m single-bore magnet, has recently been approved (in 2020) for construction at DESY. 
The projected sensitivity of Baby-IAXO is shown in Figure~\ref{fig:sensitivity_helio-LSW} as well as the possible limits with the full-size IAXO and its updated version. 
For the experiment to become sensitive in a wider axion mass range, the vacuum chamber needs to be filled with a gas in order to match the axion mass with the gas plasma frequency~\cite{article:CAST}.

%%%%%%%%%%%%%%%%%%%%%%%%%%%%%%%%%%%%%%%%%%%%%%%%%%%%%%%%%%%%%%%%%%%%%%%%%%%%%%%%%%%%%%%%%%%%%%
\section{LSW Searches}\label{sec:LSW}
The axion or ALP, if it exists, could also be produced and detected in the laboratory without relying on extraterrestrial sources.
One of the straightforward laboratory schemes is the photon regeneration, dubbed ``light shining through a wall".
Photons emitted by a strong light source are exposed in a SC dipole magnet where axions are produced via photon-axion mixing and continue to propagate through an optical wall. A similar magnet system behind the wall converts the axions back into photons with the original energy, detected by an optical sensor.
A high-finesse Fabry-Perot (PF) optical resonator is introduced in the production magnet to increase the probability of photon-to-axion conversion. 
The coherence length of the mixed state depends on the axion mass and the photon energy.
Since the axions need to first be produced and then be reconverted to photons, the process is proportional to the fourth power of the axion-to-photon coupling. 
The experimental figure of merit is given by the probability of the double conversion $a\rightarrow\gamma\rightarrow a$ as

\begin{equation}
    \mathcal{P}_{a\rightarrow\gamma\rightarrow a} = 6\times10^{-34} \left(\frac{g_{a\gamma\gamma}}{10^{-10}\,{\rm GeV^{-1}}}\right)^4 \left(\frac{B_0}{10\,T}\right)^4 \left(\frac{L}{10\,{\rm m}}\right)^4 \mathcal{F}_{P}\mathcal{F}_{R},
\end{equation}
where $\mathcal{F}_{P}$ and $\mathcal{F}_{R}$ are the enhancement factor representing power built-up by the optics systems in the production and reconversion regions, respectively.

\begin{comment}
\begin{figure}[h]
    \centering
    \includegraphics[width=1.\linewidth]{LSW_BBF_wall.png}
    \caption{Conceptual design of LSW experiments. An optical resonator is on the left side of the wall in a strong dipole magnetic field where the photons can convert to axions in each pass, while on the right side they can convert back to photons focused by the lens located just before a low dark-count photo-multiplier (adapted from  Ref.~\cite{article:BFRT1}).}
    \label{fig:LSW}
\end{figure}
\end{comment}

The first LSW experiment was performed by the BFRT (Brookhaven-Fermilab-Rochester-Trieste) Collaboration using a pair of 4.4m-long dipole magnets to search for light scalar or pseudoscalar particles for $m_a<10^{-3}$\,eV~\cite{article:BFRT1, article:BFRT2}. 
OSQAR (Optical Search of QED vacuum magnetic birefringence, Axion and photon Regeneration) 
used a buffer gas at a specific pressure to amplify the photon-axion conversions in LHC dipole magnets at CERN~\cite{article:OSQAR1}. 
The most stringent laboratory limits for $m_a<0.3$\,meV was achieved by employing 9T transverse magnetic fields over the unprecedented length of $2\times14.3$\,m~\cite{article:OSQAR3}. 
Meanwhile, the ALPS (Any Light Particle Search) Collaboration recycled several SC HERA dipole magnets to set up an experiment at the site of DESY. 
An upgraded laser system for increased power and incorporation of a low-noise high-efficiency charge-coupled device camera allowed for experimental constraints comparable to OSQAR on the existence of low mass axion-like particle~\cite{article:ALPS}.

A new breakthrough idea was proposed with an additional FP resonator built on the reconversion side~\cite{article:enhanced_LSW}. 
This second resonator is actively phase-locked together with the first FP resonator in the production magnet, significantly increasing the probability of the inverse Primakoff process in the reconversion region.
Based on this concept, the next generation of ALPS, named ALPS II, is currently under construction in a straight section of the HERA tunnel.
Using dual optical cavities installed in two strings of 12 SC dipoles, each of which has 8.8\,m length (i.e., the total length of a string is about 100\,m) and 5.3\,T field strength, the experiment aims at a substantial improvement of the current laboratory bound on $g_{a\gamma\gamma}$ by a factor $\sim10^3$ in the near future~\cite{article:ALPSII}, potentially making it more sensitive than the Sun for the first time.

%%%%%%%%%%%%%%%%%%%%%%%%%%%%%%%%%%%%%%%%%%%%%%%%%%%%%%%%%%%%%%%%%%%%%%%%%%%%%%%%%%%%%%%%%%%%%%
\section{EDM, CP-violation and Axion}\label{sec:axion_EDM}

The existence of the axion field can address an important question in particle physics; why is the neutron EDM so small? i.e., why the experimental limit is more than ten orders of magnitude smaller than anticipated from QCD.
This field dynamically cancels the $\Bar\theta$ term out of the QCD Lagrangian and hence effectively brings the EDM value of hadrons very close to zero.
However, any $CP$-violation in nature slightly shifts the EDM value away from zero due to non-perturbative QCD effects. 
With the $CP$-violating phase already observed in the electroweak (EW) interactions, represented by the CKM phase, a non-zero (albeit small) hadronic EDM is expected.
Additional $CP$-violating phases beyond the Standard Model will move its value further away. 
In fact, a new, much larger $CP$-violating source is believed to exist since the observed EW $CP$-violation is too small to account for the prevalence of matter over antimatter in the present universe; one of the biggest questions in physics today.
Therefore, once there is an EDM observed in a hadronic system, a number of other similar systems are needed to be examined to uncover the $CP$-violating source.

The storage ring method can improve the sensitivity for direct measurements of the hadronic EDM by three to four orders of magnitude over the present limits~\cite{article:Graham21, article:Chang19, article:Hempelmann17, article:Guidoboni16, article:Anastassopoulos16, article:Metodiev15, article:Metodiev14, article:Morse13, article:Brantjes12, article:Bennett09, article:Farley04}. 
The method utilizes a high-intensity beam of longitudinally-polarized protons (order $10^{11}$) stored in a circular ring with electric bending fields to accurately gauge the proton spin precession over the spin coherent time (about 1000\,s).
The proton beam is set up to be at the so-called magic momentum, along which the proton spin is aligned at all times, freezing the average spin precession on the horizontal plane~\cite{article:Haciomeroglu19}.
In such configuration, the presence of a non-zero EDM develops the spin precession on the vertical plane under the influence of the radial electric field.
To realize the concept, the beam and spin dynamics have been simulated with high precision and beam position monitoring systems have been developed using SQUID gradiometers. 
A new hybrid ring design, featured by a highly symmetric configuration of the magnetic focusing systems, was proposed to enable simultaneous storage of the clockwise and counterclockwise beams such that the systematic errors can be greatly reduced~\cite{article:Omarov20}, enabling the experiment to be performed using only currently available technology.

\begin{figure}
    \centering
    \includegraphics[width=0.7\linewidth]{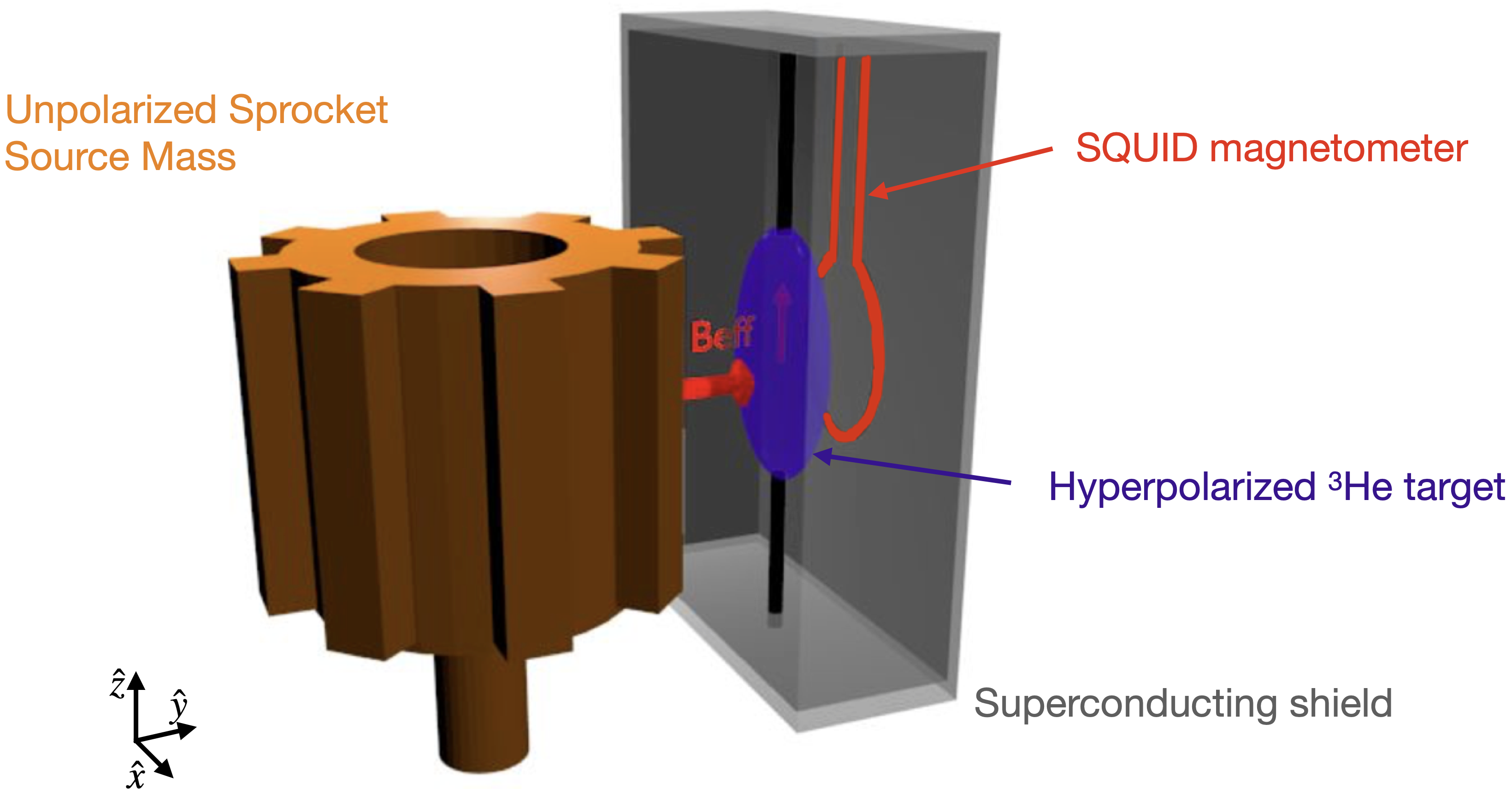}
    \caption{Schematic view of the ARIADNE setup. 
    The rotation of the unpolarized segmented cylinder sources a modulating effective magnetic field which acts on the laser polarized $^3$He gas.
    Synchronization of the modulating frequency with the Larmor precession of $^3$He will accumulate its transverse nuclear spin, which can be sensed by the SQUID magnetometer.
    }
    \label{fig:ARIADNE_exp}
\end{figure}

\begin{figure}
    \centering
    \includegraphics[width=1\linewidth]{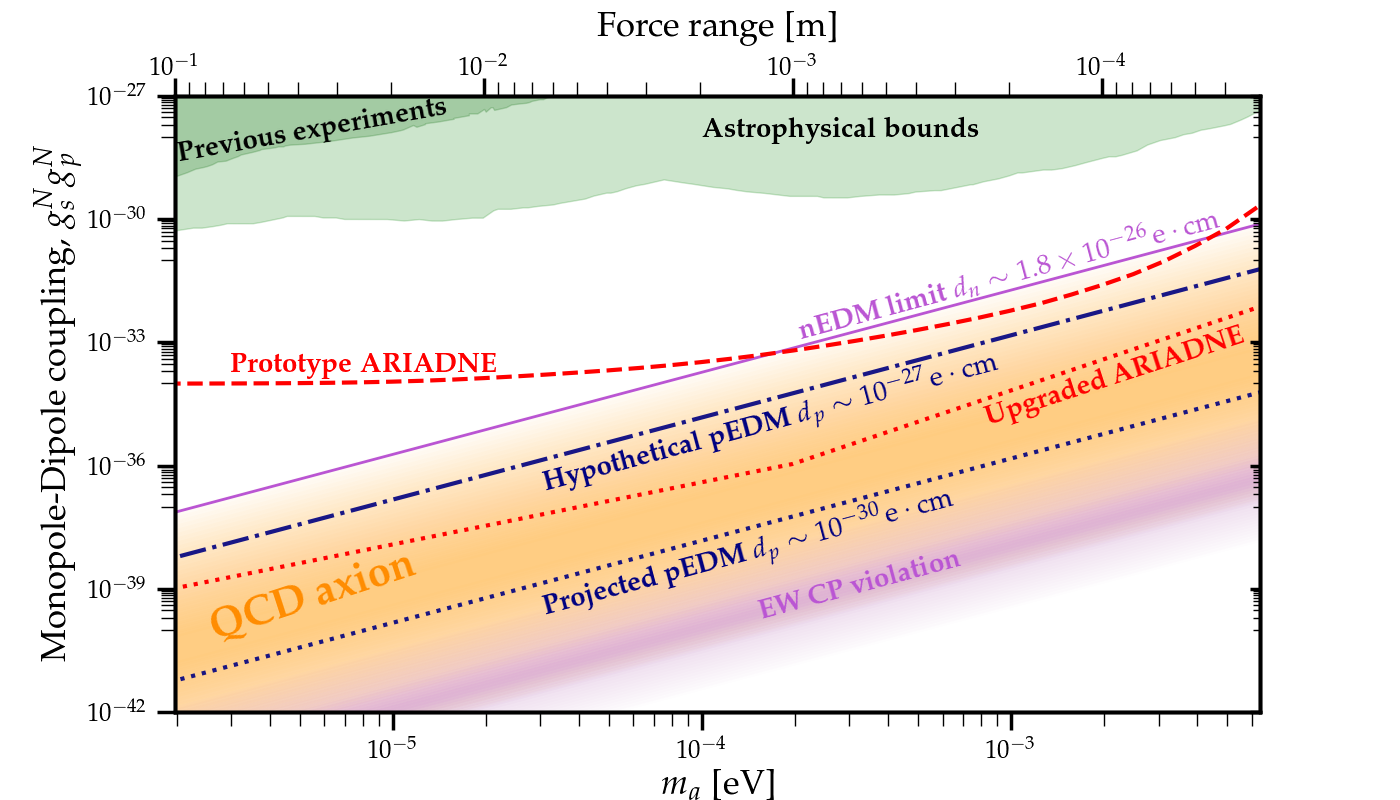}
    \caption{Future projections of the ARIADNE and storage ring proton EDM searches shown in terms nuclear monopole-dipole coupling vs. axion mass.
    The desired sensitivities by the prototype and upgraded ARIADNE experiments are represented by the dashed and dotted lines in red overlapped on the QCD axion band (in the orange area).
    The purple solid lines correspond to the current experimental upper limit on nEDM and theoretical lower bound imposed by the EW $CP$-violating phase.}
    \label{fig:ariadne_pEDM}
\end{figure}

A generic experiment searching for axion-mediated $CP$-violating forces is currently under development by an international collaboration, ARIADNE (Axion Resonant InterAction Detection Experiment)~\cite{article:ARIADNE}.
In the presence of an anomalous $CP$-violating phase, $CP$-odd axion fields can mediate short-range monopole-dipole interactions between matter objects. 
The proposal is based on the resonant coupling between the rotational frequency of a source mass and a NMR sample with a matching Larmor-spin precession frequency.
The experimental scheme, shown in Fig.~\ref{fig:ARIADNE_exp}, involves an unpolarized sprocket mass as a monopole source, polarized $^3{\rm He}$ nuclei for a dipole spin system, and a SQUID magnetometer to pick up the resonantly enhanced precession signal~\cite{article:ARIADNE_SC}.
The sensitive mass (frequency) range by this spin-dependent interaction is $0.1–10$\,meV ($25\,{\rm GHz}–2.5$\,THz), which is practically unreachable by the haloscope experiments.
The Collaboration plans to perform the experiment using hyperpolarized $^3$He gas as the NMR target with a future upgrade for ultimate sensitivity by scaling the size of the apparatus and increasing the sample density (liquid phase of $^3$He).
The projected sensitivities of the ARIADNE and storage ring EDM experiments are shown in Fig.~\ref{fig:ariadne_pEDM}.

Even though the underlying principle of the axion-mediated spin-dependent interactions does not necessarily demand the axion to be dark matter, it requires new $CP$-violating phases in order for the proposed experiment to be sensitive enough to detect a signal.
Therefore, any signal observed by ARIADNE can reveal both the axion mechanism and the existence of new $CP$-violation within its sensitive mass range.
If the physics of QCD axions is realized in nature, the ARIADNE signal is expected to lie above the SM limit set by the EW CKM $CP$-violating phase and below the current experimental limit on the neutron EDM~\cite{article:nEDM} in Fig.~\ref{fig:ariadne_pEDM}.
If there is no physics beyond the Standard Model, on the other hand, the experiment is supposed to observe a signal within the area in purple on the bottom of the figure, depending on the axion mass. 

Lastly, as mentioned earlier, most of the sensitive experiments looking for the axion signal presume that the axions solely compose the local dark matter halo.
This implies that null results do not really exclude the hypothesis of existence of this particle, but they merely rule it out as a dominant contributor to dark matter within the search range.
However, the simultaneous consideration of the experimental results between ARIADNE and nucleon EDM measurements could provide decisive information.
For example, if a proton or neutron EDM were measured to be $10^{-27}\, e \cdot {\rm cm}$  (represented by the navy dash-dotted line in Fig.~\ref{fig:ariadne_pEDM}) while the ARIADNE experiment failed to see any signal down to the ultimate sensitivity level (corresponding to ``Upgraded ARIADNE" in the same figure), then one could declare that axions do not exist in that accessible mass range.
On the other hand, a combination of the resonant signal observed by ARIADNE and vain attempts by the storage ring EDM measurements up to the projected limit (the navy dotted line) could be used to claim the existence of the axion as a new force mediator, which does not solve the strong-$CP$ problem.
In this regard, the combination of the two independent experiments would provide a unique capability to shed light on the axion physics in the related axion mass range.

%%%%%%%%%%%%%%%%%%%%%%%%%%%%%%%%%%%%%%%%%%%%%%%%%%%%%%%%%%%%%%%%%%%%%%%%%%%%%%%%%%%%%%%%%%%%%%
\section{Conclusions}
During the last few decades, substantial experimental efforts have been made to search for the invisible axion, which is believed to address two fundamental questions in physics $-$ the strong-$CP$ problem and the dark matter mystery.
Different detection schemes have been developed to probe different mass ranges and various new ideas and concepts have been  proposed based on newly emerging technologies and interdisciplinary research.
The haloscope experiments have entered a very exciting phase, reaching the theoretically interesting territory to test the favoured axion models and extending the search range to cover a large fraction of the microwave region.
In particular, the cavity haloscopes have provided the most sensitive experimental approach so far.
For instance, ADMX and HAYSTAC have approached interesting sensitivity levels for the first time within relatively low mass regions below 20\,$\mu$eV.
CAPP has been performing intensive R\&D projects in various areas and demonstrated its capability of exploring the axion physics up to 100\,$\mu$eV and even testing the QCD axions as a partial composition of the local dark matter halo.
MADMAX is expected to expand the search range up to 100\,GHz (400\,$\mu$eV) by exploiting dielectric properties.

Apart from them, a number of interesting search programs relying on lumped element circuits (e.g., DM Radio), and NMR techniques (e.g., CASPEr) seem promising to cover a broad range below the  QCD-band regions within the next few operation phases.
Solar axions can also reveal the axion physics particularly above the meV range, leading to an international proposal of a next generation large-scale helioscope, IAXO.
ALPS II is currently under upgrade to provide an independent search method with increased sensitivity of the LSW scheme for axion-like particles.
Finally, a combined interpretation of search results from the ARIADNE and nucleon EDM experiments could address the axion physics in a decisive manner particularly in the high mass regions, which are not easily reachable by any other  terrestrial experiment.
Overall, it is foreseen that the next decade is going to be very interesting and exciting in the axion field and indeed could see the discovery of the axion in any part of the parameter space.

\section*{Acknowledgements}
This work of Y.K. Semertzidis and S.W. Youn is supported by IBS-R017-D1. We thank J. Jeong and Y. Kim for drawing some of the figures in the text. We also thank F. Chadha-Day, J. Ellis, and D. J. E. Marsh for their editorial comments as well as our numerous colleagues at CAPP for their collaboration and great work, making axion dark matter research such an exciting endeavor.

\bibliographystyle{naturemag}
\bibliography{main}

\begin{thebibliography}{100}
\expandafter\ifx\csname url\endcsname\relax
  \def\url#1{\texttt{#1}}\fi
\expandafter\ifx\csname urlprefix\endcsname\relax\def\urlprefix{URL }\fi
\providecommand{\bibinfo}[2]{#2}
\providecommand{\eprint}[2][]{\url{#2}}

\bibitem{article:PQ77}
\bibinfo{author}{Peccei, R.~D.} \& \bibinfo{author}{Quinn, H.~R.}
\newblock \bibinfo{title}{{CP conservation in the presence of instantons}}.
\newblock \emph{\bibinfo{journal}{Phys. Rev. Lett.}}
  \textbf{\bibinfo{volume}{38}}, \bibinfo{pages}{1440} (\bibinfo{year}{1977}).
\newblock \urlprefix\url{https://doi.org/10.1103/PhysRevLett.38.1440}.

\bibitem{article:Weinberg78}
\bibinfo{author}{Weinberg, S.}
\newblock \bibinfo{title}{{A new light boson?}}
\newblock \emph{\bibinfo{journal}{Phys. Rev. Lett.}}
  \textbf{\bibinfo{volume}{40}}, \bibinfo{pages}{223} (\bibinfo{year}{1978}).
\newblock \urlprefix\url{https://doi.org/10.1103/PhysRevLett.40.223}.

\bibitem{article:Wilczek78}
\bibinfo{author}{Wilczek, F.}
\newblock \bibinfo{title}{{Problem of strong P and T invariance in the presence
  of instantons}}.
\newblock \emph{\bibinfo{journal}{Phys. Rev. Lett.}}
  \textbf{\bibinfo{volume}{40}}, \bibinfo{pages}{229} (\bibinfo{year}{1978}).
\newblock \urlprefix\url{https://doi.org/10.1103/PhysRevLett.40.229}.

\bibitem{article:KSVZ1}
\bibinfo{author}{Kim, J.~E.}
\newblock \bibinfo{title}{{Weak interaction singlet and strong CP invariance}}.
\newblock \emph{\bibinfo{journal}{Phys. Rev. Lett.}}
  \textbf{\bibinfo{volume}{43}}, \bibinfo{pages}{103} (\bibinfo{year}{1979}).
\newblock \urlprefix\url{https://doi.org/10.1103/PhysRevLett.43.103}.

\bibitem{article:KSVZ2}
\bibinfo{author}{Shifman, M.~A.}, \bibinfo{author}{Vainshtein, A.~I.} \&
  \bibinfo{author}{Zakharov, V.~I.}
\newblock \bibinfo{title}{{Can confinement ensure natural CP invariance of
  strong interactions?}}
\newblock \emph{\bibinfo{journal}{Nucl. Phys. B}}
  \textbf{\bibinfo{volume}{166}}, \bibinfo{pages}{4933} (\bibinfo{year}{1980}).
\newblock \urlprefix\url{https://doi.org/10.1016/0550-3213(80)90209-6}.

\bibitem{article:DFSZ1}
\bibinfo{author}{Zhitnitsky, A.~P.}
\newblock \bibinfo{title}{{On possible suppression of the axion hadron
  interactions}}.
\newblock \emph{\bibinfo{journal}{Sov. J. Nucl. Phys.}}
  \textbf{\bibinfo{volume}{31}}, \bibinfo{pages}{260} (\bibinfo{year}{1980}).

\bibitem{article:DFSZ2}
\bibinfo{author}{Dine, M.}, \bibinfo{author}{Fischler, W.} \&
  \bibinfo{author}{Srednicki, M.}
\newblock \bibinfo{title}{{A simple solution to the strong CP problem with a
  harmless axion}}.
\newblock \emph{\bibinfo{journal}{Phys. Lett. B}}
  \textbf{\bibinfo{volume}{104}}, \bibinfo{pages}{199} (\bibinfo{year}{1981}).
\newblock \urlprefix\url{https://doi.org/10.1016/0370-2693(81)90590-6}.

\bibitem{article:string1}
\bibinfo{author}{Svrcek, P.} \& \bibinfo{author}{Witten, E.}
\newblock \bibinfo{title}{{Axions in string theory}}.
\newblock \emph{\bibinfo{journal}{J. of High Energ. Phys.}}
  \textbf{\bibinfo{volume}{2006}}, \bibinfo{pages}{051} (\bibinfo{year}{2006}).
\newblock \urlprefix\url{https://doi.org/10.1088/1126-6708/2006/06/051}.

\bibitem{article:string2}
\bibinfo{author}{Arvanitaki, A.}, \bibinfo{author}{Dimopoulos, S.},
  \bibinfo{author}{Dubovsky, S.}, \bibinfo{author}{Kaloper, N.} \&
  \bibinfo{author}{March-Russell, J.}
\newblock \bibinfo{title}{{String axiverse}}.
\newblock \emph{\bibinfo{journal}{Phys. Rev. D}} \textbf{\bibinfo{volume}{81}},
  \bibinfo{pages}{123530} (\bibinfo{year}{2010}).
\newblock \urlprefix\url{https://doi.org/10.1103/PhysRevD.81.123530}.

\bibitem{article:string3}
\bibinfo{author}{Ringwald, A.}
\newblock \bibinfo{title}{{Axions and Axion-Like Particles}}
  (\bibinfo{year}{2014}).
\newblock \eprint{1407.0546}.

\bibitem{article:axion_cosmology1}
\bibinfo{author}{Preskill, J.}, \bibinfo{author}{Wise, M.~B.} \&
  \bibinfo{author}{Wilczek, F.}
\newblock \bibinfo{title}{{Cosmology of the invisible axion}}.
\newblock \emph{\bibinfo{journal}{Phys. Lett. B}}
  \textbf{\bibinfo{volume}{120}}, \bibinfo{pages}{127} (\bibinfo{year}{1983}).
\newblock \urlprefix\url{https://doi.org/10.1016/0370-2693(83)90637-8}.

\bibitem{article:axion_cosmology2}
\bibinfo{author}{Abbott, L.~F.} \& \bibinfo{author}{Sikivie, P.}
\newblock \bibinfo{title}{{A cosmological bound on the invisible axion}}.
\newblock \emph{\bibinfo{journal}{Phys. Lett. B}}
  \textbf{\bibinfo{volume}{120}}, \bibinfo{pages}{133} (\bibinfo{year}{1983}).
\newblock \urlprefix\url{https://doi.org/10.1016/0370-2693(83)90638-X}.

\bibitem{article:axion_cosmology3}
\bibinfo{author}{Dine, M.} \& \bibinfo{author}{Fischler, W.}
\newblock \bibinfo{title}{{The not-so-harmless axion}}.
\newblock \emph{\bibinfo{journal}{Phys. Lett. B}}
  \textbf{\bibinfo{volume}{120}}, \bibinfo{pages}{137} (\bibinfo{year}{1983}).
\newblock \urlprefix\url{https://doi.org/10.1016/0370-2693(83)90639-1}.

\bibitem{article:SN1987A}
\bibinfo{author}{Chang, J.~H.}, \bibinfo{author}{Essig, R.} \&
  \bibinfo{author}{McDermott, S.~D.}
\newblock \bibinfo{title}{{Supernova 1987A constraints on Sub-GeV dark sectors,
  millicharged particles, the QCD axion, and an axion-like particle}}.
\newblock \emph{\bibinfo{journal}{J. High Energy Phys.}}
  \textbf{\bibinfo{volume}{09}}, \bibinfo{pages}{051} (\bibinfo{year}{2018}).
\newblock \urlprefix\url{https://doi.org/10.1007/JHEP09(2018)051}.

\bibitem{article:Turner}
\bibinfo{author}{Turner, M.~S.}
\newblock \bibinfo{title}{{Periodic signatures for the detection of cosmic
  axions}}.
\newblock \emph{\bibinfo{journal}{Phys. Rev. D}} \textbf{\bibinfo{volume}{42}},
  \bibinfo{pages}{3572} (\bibinfo{year}{1990}).
\newblock \urlprefix\url{https://doi.org/10.1103/PhysRevD.42.3572}.

\bibitem{article:RBF1}
\bibinfo{author}{DePanfilis, S.} \emph{et~al.}
\newblock \bibinfo{title}{{Limits on the abundance and coupling of cosmic
  axions at $4.5<m_a<5.0\,\mu$eV}}.
\newblock \emph{\bibinfo{journal}{Phys. Rev. Lett.}}
  \textbf{\bibinfo{volume}{59}}, \bibinfo{pages}{839} (\bibinfo{year}{1987}).
\newblock \urlprefix\url{https://doi.org/10.1103/PhysRevLett.59.839}.

\bibitem{article:ADMX}
\bibinfo{author}{Hagmann, C.} \emph{et~al.}
\newblock \bibinfo{title}{{Results from a High-Sensitivity Search for Cosmic
  Axions}}.
\newblock \emph{\bibinfo{journal}{Phys. Rev. Lett.}}
  \textbf{\bibinfo{volume}{80}}, \bibinfo{pages}{2043} (\bibinfo{year}{1998}).
\newblock \urlprefix\url{https://doi.org/10.1103/PhysRevLett.80.2043}.

\bibitem{article:HAYSTAC1}
\bibinfo{author}{Brubaker, B.~M.} \emph{et~al.}
\newblock \bibinfo{title}{{First Results from a Microwave Cavity Axion Search
  at 24\,$\mu$eV}}.
\newblock \emph{\bibinfo{journal}{Phys. Rev. Lett.}}
  \textbf{\bibinfo{volume}{118}}, \bibinfo{pages}{061302}
  (\bibinfo{year}{2017}).
\newblock \urlprefix\url{https://doi.org/10.1103/PhysRevLett.118.061302}.

\bibitem{article:Yannis19}
\bibinfo{author}{Semertzidis, Y.~K.} \emph{et~al.}
\newblock \bibinfo{title}{{Axion Dark Matter Research with IBS/CAPP}}
  (\bibinfo{year}{2019}).
\newblock \eprint{1910.11591}.

\bibitem{article:GNOME}
\bibinfo{author}{Pospelov, M.} \emph{et~al.}
\newblock \bibinfo{title}{{Detecting Domain Walls of Axionlike Models Using
  Terrestrial Experiments}}.
\newblock \emph{\bibinfo{journal}{Phys. Rev. Lett.}}
  \textbf{\bibinfo{volume}{110}}, \bibinfo{pages}{021803}
  (\bibinfo{year}{2013}).
\newblock \urlprefix\url{https://doi.org/10.1103/PhysRevLett.110.021803}.

\bibitem{proc:Caspers}
\bibinfo{author}{Caspers, F.} \& \bibinfo{author}{Semertzidis, Y.}
\newblock \bibinfo{title}{Ferri-magnetic resonance, magnetostatic waves and
  open resonators for axion detection}.
\newblock In \bibinfo{editor}{C., J.} \& \bibinfo{editor}{A., M.} (eds.)
  \emph{\bibinfo{booktitle}{Proc. of the Workshop on Cosmic Axions}},
  \bibinfo{pages}{173} (\bibinfo{publisher}{World Scientific Pub. Co.},
  \bibinfo{address}{Singapore}, \bibinfo{year}{1990}).

\bibitem{article:ARIADNE}
\bibinfo{author}{Arvanitaki, A.} \& \bibinfo{author}{Geraci, A.~A.}
\newblock \bibinfo{title}{{Resonantly Detecting Axion-Mediated Forces with
  Nuclear Magnetic Resonance}}.
\newblock \emph{\bibinfo{journal}{Phys. Rev. Lett.}}
  \textbf{\bibinfo{volume}{113}}, \bibinfo{pages}{161801}
  (\bibinfo{year}{2014}).
\newblock \urlprefix\url{https://doi.org/10.1103/PhysRevLett.113.161801}.

\bibitem{article:Graham21}
\bibinfo{author}{Graham, P.~W.} \emph{et~al.}
\newblock \bibinfo{title}{{Storage ring probes of dark matter and dark
  energy}}.
\newblock \emph{\bibinfo{journal}{Phys. Rev. D}}
  \textbf{\bibinfo{volume}{103}}, \bibinfo{pages}{055010}
  (\bibinfo{year}{2021}).
\newblock \urlprefix\url{https://doi.org/10.1103/PhysRevD.103.055010}.

\bibitem{article:CASPEr}
\bibinfo{author}{Budker, D.}, \bibinfo{author}{Graham, P.~W.},
  \bibinfo{author}{Ledbetter, M.}, \bibinfo{author}{Rajendran, S.} \&
  \bibinfo{author}{Sushkov, A.~O.}
\newblock \bibinfo{title}{{Proposal for a Cosmic Axion Spin Precession
  Experiment (CASPEr)}}.
\newblock \emph{\bibinfo{journal}{Phys. Rev. X}} \textbf{\bibinfo{volume}{4}},
  \bibinfo{pages}{021030} (\bibinfo{year}{2014}).
\newblock \urlprefix\url{https://doi.org/10.1103/PhysRevX.4.021030}.

\bibitem{article:Chang19}
\bibinfo{author}{Chang, S.~P.} \emph{et~al.}
\newblock \bibinfo{title}{{Axionlike dark matter search using the storage ring
  EDM method}}.
\newblock \emph{\bibinfo{journal}{Phys. Rev. D}} \textbf{\bibinfo{volume}{99}},
  \bibinfo{pages}{083002} (\bibinfo{year}{2019}).
\newblock \urlprefix\url{https://doi.org/10.1103/PhysRevD.99.083002}.

\bibitem{article:Brun19}
\bibinfo{author}{Brun, P.} \emph{et~al.}
\newblock \bibinfo{title}{{A new experimental approach to probe QCD axion dark
  matter in the mass range above $40\,{\mu \rm eV}$}}.
\newblock \emph{\bibinfo{journal}{Eur. Phys. J. C}}
  \textbf{\bibinfo{volume}{79}}, \bibinfo{pages}{186} (\bibinfo{year}{2019}).
\newblock \urlprefix\url{https://doi.org/10.1140/epjc/s10052-019-6683-x}.

\bibitem{article:new_force}
\bibinfo{author}{Moody, J.~E.} \& \bibinfo{author}{Wilczek, F.}
\newblock \bibinfo{title}{{New macroscopic forces?}}
\newblock \emph{\bibinfo{journal}{Phys. Rev. D}} \textbf{\bibinfo{volume}{30}},
  \bibinfo{pages}{130} (\bibinfo{year}{1984}).
\newblock \urlprefix\url{https://doi.org/10.1103/PhysRevD.30.130}.

\bibitem{article:non_Newtonian_force}
\bibinfo{author}{Geraci, A.~A.}, \bibinfo{author}{Smullin, S.~J.},
  \bibinfo{author}{Weld, D.~M.}, \bibinfo{author}{Chiaverini, J.} \&
  \bibinfo{author}{Kapitulnik, A.}
\newblock \bibinfo{title}{{Improved constraints on non-Newtonian forces at 10
  microns}}.
\newblock \emph{\bibinfo{journal}{Phys. Rev. D}} \textbf{\bibinfo{volume}{78}},
  \bibinfo{pages}{022002} (\bibinfo{year}{2008}).
\newblock \urlprefix\url{https://doi.org/10.1103/PhysRevD.78.022002}.

\bibitem{article:Sikivie83}
\bibinfo{author}{Sikivie, P.}
\newblock \bibinfo{title}{{Experimental Tests of the Invisible Axion}}.
\newblock \emph{\bibinfo{journal}{Phys. Rev. Lett.}}
  \textbf{\bibinfo{volume}{51}}, \bibinfo{pages}{1415} (\bibinfo{year}{1983}).
\newblock \urlprefix\url{https://doi.org/10.1103/PhysRevLett.51.1415}.

\bibitem{article:P&P}
\bibinfo{author}{Pshirkov, M.~S.} \& \bibinfo{author}{Popov, S.~B.}
\newblock \bibinfo{title}{{Conversion of Dark matter axions to photons in
  magnetospheres of neutron stars}}.
\newblock \emph{\bibinfo{journal}{J. Exp. Theor. Phys.}}
  \textbf{\bibinfo{volume}{108}}, \bibinfo{pages}{384} (\bibinfo{year}{2009}).
\newblock \urlprefix\url{https://doi.org/10.1134/S1063776109030030}.

\bibitem{article:Huang}
\bibinfo{author}{F.~P.~Huang, T.~S., K.~Kadota} \& \bibinfo{author}{Tashiro,
  H.}
\newblock \bibinfo{title}{{Radio telescope search for the resonant conversion
  of cold dark matter axions from the magnetized astrophysical sources}}.
\newblock \emph{\bibinfo{journal}{Phys. Rev. D}} \textbf{\bibinfo{volume}{97}},
  \bibinfo{pages}{123001} (\bibinfo{year}{2018}).
\newblock \urlprefix\url{https://doi.org/10.1103/PhysRevD.97.123001}.

\bibitem{article:Hook}
\bibinfo{author}{Anson~Hook, B. R.~S., Yonatan~Kahn} \& \bibinfo{author}{Sun,
  Z.}
\newblock \bibinfo{title}{{Radio Signals from Axion Dark Matter Conversion in
  Neutron Star Magnetospheres}}.
\newblock \emph{\bibinfo{journal}{Phys. Rev. Lett.}}
  \textbf{\bibinfo{volume}{121}}, \bibinfo{pages}{241102}
  (\bibinfo{year}{2018}).
\newblock \urlprefix\url{https://doi.org/10.1103/PhysRevLett.121.241102}.

\bibitem{article:Safdi}
\bibinfo{author}{B.~R.~Safdi, Z.~S.} \& \bibinfo{author}{Chen, A.~Y.}
\newblock \bibinfo{title}{{Detecting Axion Dark Matter with Radio Lines from
  Neutron Star Populations}}.
\newblock \emph{\bibinfo{journal}{Phys. Rev. D}} \textbf{\bibinfo{volume}{99}},
  \bibinfo{pages}{123021} (\bibinfo{year}{2019}).
\newblock \urlprefix\url{https://doi.org/10.1103/PhysRevD.99.123021}.

\bibitem{article:Leroy}
\bibinfo{author}{M.~Leroy, T. D. P.~E., M.~Chianese} \&
  \bibinfo{author}{Weniger, C.}
\newblock \bibinfo{title}{{Radio Signal of Axion-Photon Conversion in Neutron
  Stars: A Ray Tracing Analysis}}.
\newblock \emph{\bibinfo{journal}{Phys. Rev. D}}
  \textbf{\bibinfo{volume}{101}}, \bibinfo{pages}{123003}
  (\bibinfo{year}{2020}).
\newblock \urlprefix\url{https://doi.org/10.1103/PhysRevD.101.123003}.

\bibitem{article:Battye}
\bibinfo{author}{R.~A.~Battye, J. I. M. F.~P., B.~Garbrecht} \&
  \bibinfo{author}{Srinivasan, S.}
\newblock \bibinfo{title}{{Dark matter axion detection in the
  radio/mm-waveband}}.
\newblock \emph{\bibinfo{journal}{Phys. Rev. D}}
  \textbf{\bibinfo{volume}{102}}, \bibinfo{pages}{023504}
  (\bibinfo{year}{2020}).
\newblock \urlprefix\url{https://doi.org/10.1103/PhysRevD.102.023504}.

\bibitem{article:PSR1}
\bibinfo{author}{Darling, J.}
\newblock \bibinfo{title}{{Search for Axionic Dark Matter Using the Magnetar
  PSR J1745-2900}}.
\newblock \emph{\bibinfo{journal}{Phys. Rev. Lett.}}
  \textbf{\bibinfo{volume}{125}}, \bibinfo{pages}{121103}
  (\bibinfo{year}{2020}).
\newblock \urlprefix\url{https://doi.org/10.1103/PhysRevLett.125.121103}.

\bibitem{article:PSR2}
\bibinfo{author}{Darling, J.}
\newblock \bibinfo{title}{{New Limits on Axionic Dark Matter from the Magnetar
  PSR J1745-2900}}.
\newblock \emph{\bibinfo{journal}{Astrophys. J. Lett.}}
  \textbf{\bibinfo{volume}{900}}, \bibinfo{pages}{L28} (\bibinfo{year}{2020}).
\newblock \urlprefix\url{https://doi.org/10.3847/2041-8213/abb23f}.

\bibitem{article:Foster}
\bibinfo{author}{Foster, J.~W.} \emph{et~al.}
\newblock \bibinfo{title}{{Green Bank and Effelsberg Radio Telescope Searches
  for Axion Dark Matter Conversion in Neutron Star Magnetospheres}}.
\newblock \emph{\bibinfo{journal}{Phys. Rev. Lett.}}
  \textbf{\bibinfo{volume}{125}}, \bibinfo{pages}{171301}
  (\bibinfo{year}{2020}).
\newblock \urlprefix\url{https://doi.org/10.1103/PhysRevLett.125.171301}.

\bibitem{article:Chen}
\bibinfo{author}{A.~Y.~Chen, F.~C.} \& \bibinfo{author}{Spitkovsky, A.}
\newblock \bibinfo{title}{{Filling the magnetospheres of weak pulsars}}.
\newblock \emph{\bibinfo{journal}{Astrophys. J}}
  \textbf{\bibinfo{volume}{889}}, \bibinfo{pages}{69} (\bibinfo{year}{2020}).
\newblock \urlprefix\url{https://doi.org/10.3847/1538-4357/ab5c20}.

\bibitem{article:Carrasco}
\bibinfo{author}{Carrasco, F.} \& \bibinfo{author}{Shibata, M.}
\newblock \bibinfo{title}{{Magnetosphere of an orbit- ing neutron star}}.
\newblock \emph{\bibinfo{journal}{Phys. Rev. D}}
  \textbf{\bibinfo{volume}{101}}, \bibinfo{pages}{063017}
  (\bibinfo{year}{2020}).
\newblock \urlprefix\url{https://doi.org/10.1103/PhysRevD.101.063017}.

\bibitem{article:Witte}
\bibinfo{author}{Samuel J.~Witte, T. D. P.~E., Dion~Noordhuis} \&
  \bibinfo{author}{Weniger, C.}
\newblock \bibinfo{title}{{Axion-Photon Conversion in Neutron Star
  Magnetospheres: The Role of the Plasma in the Goldreich-Julian Model}}
  (\bibinfo{year}{2021}).
\newblock \eprint{2104.07670}.

\bibitem{article:Sikivie21}
\bibinfo{author}{Sikivie, P.}
\newblock \bibinfo{title}{{Invisible axion search methods}}.
\newblock \emph{\bibinfo{journal}{Rev. Mod. Phys.}}
  \textbf{\bibinfo{volume}{93}}, \bibinfo{pages}{015004}
  (\bibinfo{year}{2021}).
\newblock \urlprefix\url{https://doi.org/10.1103/RevModPhys.93.015004}.

\bibitem{article:APPEC}
\bibinfo{author}{Billard, J.} \emph{et~al.}
\newblock \bibinfo{title}{{Direct Detection of Dark Matter -- APPEC Committee
  Report}}  (\bibinfo{year}{2021}).
\newblock \eprint{2104.07634}.

\bibitem{article:RBF2}
\bibinfo{author}{Wuensch, W.} \emph{et~al.}
\newblock \bibinfo{title}{{Results of a laboratory search for cosmic axions and
  other weakly coupled light particles}}.
\newblock \emph{\bibinfo{journal}{Phys. Rev. D}} \textbf{\bibinfo{volume}{40}},
  \bibinfo{pages}{3153} (\bibinfo{year}{1989}).
\newblock \urlprefix\url{https://doi.org/10.1103/PhysRevD.40.3153}.

\bibitem{article:UF}
\bibinfo{author}{Hagmann, C.} \emph{et~al.}
\newblock \bibinfo{title}{{Results from a search for cosmic axionss}}.
\newblock \emph{\bibinfo{journal}{Phys. Rev. D}} \textbf{\bibinfo{volume}{42}},
  \bibinfo{pages}{1297} (\bibinfo{year}{1990}).
\newblock \urlprefix\url{https://doi.org/10.1103/PhysRevD.42.1297}.

\bibitem{article:ADMX_KSVZ1}
\bibinfo{author}{Hagmann, C.} \emph{et~al.}
\newblock \bibinfo{title}{{Results from a High-Sensitivity Search for Cosmic
  Axions}}.
\newblock \emph{\bibinfo{journal}{Phys. Rev. Lett.}}
  \textbf{\bibinfo{volume}{80}}, \bibinfo{pages}{2043} (\bibinfo{year}{1998}).
\newblock \urlprefix\url{https://doi.org/10.1103/PhysRevLett.80.2043}.

\bibitem{article:ADMX_KSVZ2}
\bibinfo{author}{Asztalos, S.~J.} \emph{et~al.}
\newblock \bibinfo{title}{{Improved rf cavity search for halo axions}}.
\newblock \emph{\bibinfo{journal}{Phys. Rev. D}} \textbf{\bibinfo{volume}{69}},
  \bibinfo{pages}{011101 (R)} (\bibinfo{year}{2004}).
\newblock \urlprefix\url{https://doi.org/10.1103/PhysRevD.69.011101}.

\bibitem{article:ADMX_SQUID}
\bibinfo{author}{Asztalos, S.~J.} \emph{et~al.}
\newblock \bibinfo{title}{{SQUID-Based Microwave Cavity Search for Dark-Matter
  Axions}}.
\newblock \emph{\bibinfo{journal}{Phys. Rev. Lett.}}
  \textbf{\bibinfo{volume}{104}}, \bibinfo{pages}{041301}
  (\bibinfo{year}{2010}).
\newblock \urlprefix\url{https://doi.org/10.1103/PhysRevLett.104.041301}.

\bibitem{article:ADMX_Run1A}
\bibinfo{author}{Du, N.} \emph{et~al.}
\newblock \bibinfo{title}{{Search for Invisible Axion Dark Matter with the
  Axion Dark Matter Experiment}}.
\newblock \emph{\bibinfo{journal}{Phys. Rev. Lett.}}
  \textbf{\bibinfo{volume}{120}}, \bibinfo{pages}{151301}
  (\bibinfo{year}{2018}).
\newblock \urlprefix\url{https://doi.org/10.1103/PhysRevLett.120.151301}.

\bibitem{article:ADMX_Run1B}
\bibinfo{author}{Braine, T.} \emph{et~al.}
\newblock \bibinfo{title}{{Extended Search for the Invisible Axion with the
  Axion Dark Matter Experiment}}.
\newblock \emph{\bibinfo{journal}{Phys. Rev. Lett.}}
  \textbf{\bibinfo{volume}{124}}, \bibinfo{pages}{101303}
  (\bibinfo{year}{2020}).
\newblock \urlprefix\url{https://doi.org/10.1103/PhysRevLett.124.101303}.

\bibitem{article:ADMX_sidecar}
\bibinfo{author}{Boutan, C.} \emph{et~al.}
\newblock \bibinfo{title}{{Piezoelectrically Tuned Multimode Cavity Search for
  Axion Dark Matter}}.
\newblock \emph{\bibinfo{journal}{Phys. Rev. Lett.}}
  \textbf{\bibinfo{volume}{121}}, \bibinfo{pages}{261302}
  (\bibinfo{year}{2018}).
\newblock \urlprefix\url{https://doi.org/10.1103/PhysRevLett.121.261302}.

\bibitem{thesis:multicavity}
\bibinfo{author}{Kinion, D.}
\newblock \emph{\bibinfo{title}{{First Results from a Multiple-Microwave Cavity
  Search for Dark-Matter Axions}}}.
\newblock Ph.D. thesis, \bibinfo{school}{University of California Davis}
  (\bibinfo{year}{2001}).

\bibitem{article:multicavity}
\bibinfo{author}{Jeong, J.} \emph{et~al.}
\newblock \bibinfo{title}{{Phase-matching of multiple-cavity detectors for dark
  matter axion search}}.
\newblock \emph{\bibinfo{journal}{Astropart. Phys.}}
  \textbf{\bibinfo{volume}{97}}, \bibinfo{pages}{33} (\bibinfo{year}{2018}).
\newblock \urlprefix\url{https://doi.org/10.1016/j.astropartphys.2017.10.012}.

\bibitem{article:ADMX_SPD}
\bibinfo{author}{Dixit, A.~V.} \emph{et~al.}
\newblock \bibinfo{title}{{Searching for Dark Matter with a Superconducting
  Qubit}}.
\newblock \emph{\bibinfo{journal}{Phys. Rev. Lett.}}
  \textbf{\bibinfo{volume}{126}}, \bibinfo{pages}{141302}
  (\bibinfo{year}{2021}).
\newblock \urlprefix\url{https://doi.org/10.1103/PhysRevLett.126.141302}.

\bibitem{article:HAYSTAC2}
\bibinfo{author}{Zhong, L.} \emph{et~al.}
\newblock \bibinfo{title}{{Results from phase 1 of the HAYSTAC microwave cavity
  axion experiment}}.
\newblock \emph{\bibinfo{journal}{Phys. Rev. D}} \textbf{\bibinfo{volume}{97}},
  \bibinfo{pages}{092001} (\bibinfo{year}{2018}).
\newblock \urlprefix\url{https://doi.org/10.1103/PhysRevD.97.092001}.

\bibitem{article:squeezing}
\bibinfo{author}{Backes, K.~M.} \emph{et~al.}
\newblock \bibinfo{title}{{A quantum enhanced search for dark matter axions}}.
\newblock \emph{\bibinfo{journal}{Nature}} \textbf{\bibinfo{volume}{590}},
  \bibinfo{pages}{238} (\bibinfo{year}{2021}).
\newblock \urlprefix\url{https://doi.org/10.1038/s41586-021-03226-7}.

\bibitem{article:25T_magnet}
\bibinfo{author}{Gupta, R.} \emph{et~al.}
\newblock \bibinfo{title}{{Status of the 25\,T, 100\,mm Bore HTS Solenoid for
  an Axion Dark Matter Search Experiment}}.
\newblock \emph{\bibinfo{journal}{IEEE Trans. Appl. Supercond.}}
  \textbf{\bibinfo{volume}{29}}, \bibinfo{pages}{4602105}
  (\bibinfo{year}{2019}).
\newblock \urlprefix\url{https://doi.org/10.1109/TASC.2019.2902319}.

\bibitem{article:pizza_cavity}
\bibinfo{author}{Jeong, J.}, \bibinfo{author}{Youn, S.}, \bibinfo{author}{Ahn,
  S.}, \bibinfo{author}{Kim, J.~E.} \& \bibinfo{author}{Semertzidis, Y.~K.}
\newblock \bibinfo{title}{{Concept of multiple-cell cavity for axion dark
  matter search}}.
\newblock \emph{\bibinfo{journal}{Phys. Lett. B}}
  \textbf{\bibinfo{volume}{777}}, \bibinfo{pages}{412} (\bibinfo{year}{2018}).
\newblock \urlprefix\url{https://doi.org/10.1016/j.physletb.2017.12.066}.

\bibitem{article:SC_cavity}
\bibinfo{author}{Ahn, D.} \emph{et~al.}
\newblock \bibinfo{title}{{First prototype of a biaxially textured
  YBa$_2$Cu$_3$O$_{7-x}$ microwave cavity in a high magnetic field for dark
  matter axion search }}  (\bibinfo{year}{2021}).
\newblock \eprint{2103.14515}.

\bibitem{article:CAPP_JPA}
\bibinfo{author}{Kutlu, C.} \emph{et~al.}
\newblock \bibinfo{title}{{Characterization of a flux-driven Josephson
  parametric amplifier with near quantum-limited added noise for axion search
  experiments}}.
\newblock \emph{\bibinfo{journal}{Supercond. Sci. Technol.}}
  \textbf{\bibinfo{volume}{34}}, \bibinfo{pages}{085013}
  (\bibinfo{year}{2021}).
\newblock \urlprefix\url{https://doi.org/10.1088/1361-6668/abf23b}.

\bibitem{article:CAPP-8TB}
\bibinfo{author}{Lee, S.}, \bibinfo{author}{Ahn, S.}, \bibinfo{author}{Choi,
  J.}, \bibinfo{author}{Ko, B.~R.} \& \bibinfo{author}{Semertzidis, Y.~K.}
\newblock \bibinfo{title}{{Axion Dark Matter Search around 6.7\,$\mu$eV}}.
\newblock \emph{\bibinfo{journal}{Phys. Rev. Lett.}}
  \textbf{\bibinfo{volume}{124}}, \bibinfo{pages}{Phys. Rev. Lett.}
  (\bibinfo{year}{2020}).
\newblock \urlprefix\url{https://doi.org/10.1103/PhysRevLett.124.101802}.

\bibitem{article:CAPP-MC}
\bibinfo{author}{Jeong, J.} \emph{et~al.}
\newblock \bibinfo{title}{{Search for Invisible Axion Dark Matter with a
  Multiple-Cell Haloscope}}.
\newblock \emph{\bibinfo{journal}{Phys. Rev. Lett.}}
  \textbf{\bibinfo{volume}{125}}, \bibinfo{pages}{221302}
  (\bibinfo{year}{2020}).
\newblock \urlprefix\url{https://doi.org/10.1103/PhysRevLett.125.221302}.

\bibitem{article:CAPP-PACE}
\bibinfo{author}{Kwon, O.} \emph{et~al.}
\newblock \bibinfo{title}{{First Results from Axion Haloscope at CAPP around
  10.7\,$\mu$eV}}.
\newblock \emph{\bibinfo{journal}{Phys. Rev. Lett.}}
  \textbf{\bibinfo{volume}{126}}, \bibinfo{pages}{191802}
  (\bibinfo{year}{2021}).
\newblock \urlprefix\url{https://doi.org/10.1103/PhysRevLett.126.191802}.

\bibitem{article:QUAX_SCC}
\bibinfo{author}{Alesini, D.} \emph{et~al.}
\newblock \bibinfo{title}{{Galactic axions search with a superconducting
  resonant cavity}}.
\newblock \emph{\bibinfo{journal}{Phys. Rev. D}} \textbf{\bibinfo{volume}{99}},
  \bibinfo{pages}{101101(R)} (\bibinfo{year}{2019}).
\newblock \urlprefix\url{https://doi.org/10.1103/PhysRevD.99.101101}.

\bibitem{article:QUAX_SPD}
\bibinfo{author}{Kuzmin, L.~S.} \emph{et~al.}
\newblock \bibinfo{title}{{Single Photon Counter Based on a Josephson Junction
  at 14\,GHz for Searching Galactic Axions}}.
\newblock \emph{\bibinfo{journal}{IEEE Trans. Appl. Supercond.}}
  \textbf{\bibinfo{volume}{28}}, \bibinfo{pages}{7} (\bibinfo{year}{2018}).
\newblock \urlprefix\url{https://doi.org/10.1109/TASC.2018.2850019}.

\bibitem{article:ORGAN}
\bibinfo{author}{McAllister, B.~T.}
\newblock \bibinfo{title}{{The ORGAN experiment: An axion haloscope above
  15\,GHz}}.
\newblock \emph{\bibinfo{journal}{Phys. Dark Univ.}}
  \textbf{\bibinfo{volume}{18}}, \bibinfo{pages}{67} (\bibinfo{year}{2017}).
\newblock \urlprefix\url{https://doi.org/10.1016/j.dark.2017.09.010}.

\bibitem{proc:CAST-CAPP}
\bibinfo{author}{Miceli, L.}
\newblock \bibinfo{title}{Haloscope axion searches with the cast dipole magnet:
  The cast-capp/ibs detector}.
\newblock In \emph{\bibinfo{booktitle}{Proc. of the Patras Workshop on Axions,
  WIMPs and WISPs}} (\bibinfo{year}{2015}).
\newblock
  \urlprefix\url{http://dx.doi.org/10.3204/DESY-PROC-2015-02/miceli_lino}.

\bibitem{article:RADES}
\bibinfo{author}{Melc\'{o}n, A.~A.} \emph{et~al.}
\newblock \bibinfo{title}{{Axion searches with microwave filters: the RADES
  project}}.
\newblock \emph{\bibinfo{journal}{J. Cosmol. Astropart. Phys.}}
  \textbf{\bibinfo{volume}{05}}, \bibinfo{pages}{040} (\bibinfo{year}{2018}).
\newblock \urlprefix\url{https://doi.org/10.1088/1475-7516/2018/05/040}.

\bibitem{article:Orpheus}
\bibinfo{author}{Rybka, G.} \emph{et~al.}
\newblock \bibinfo{title}{{Search for dark matter axions with the Orpheus
  experiment}}.
\newblock \emph{\bibinfo{journal}{Phys. Rev. D}} \textbf{\bibinfo{volume}{91}},
  \bibinfo{pages}{011701(R)} (\bibinfo{year}{2015}).
\newblock \urlprefix\url{https://doi.org/10.1103/PhysRevD.91.011701}.

\bibitem{article:MADMAX}
\bibinfo{author}{Caldwell, A.} \emph{et~al.}
\newblock \bibinfo{title}{{Dielectric Haloscopes: A New Way to Detect Axion
  Dark Matter}}.
\newblock \emph{\bibinfo{journal}{Phys. Rev. Lett.}}
  \textbf{\bibinfo{volume}{118}}, \bibinfo{pages}{091801}
  (\bibinfo{year}{2017}).
\newblock \urlprefix\url{https://doi.org/10.1103/PhysRevLett.118.091801}.

\bibitem{article:booster}
\bibinfo{author}{Egge, J.} \emph{et~al.}
\newblock \bibinfo{title}{{A first proof of principle booster setup for the
  MADMAX dielectric haloscope}}.
\newblock \emph{\bibinfo{journal}{Eur. Phys. J. C}}
  \textbf{\bibinfo{volume}{80}}, \bibinfo{pages}{392} (\bibinfo{year}{2020}).
\newblock \urlprefix\url{https://doi.org/10.1140/epjc/s10052-020-7985-8}.

\bibitem{article:MADMAX-3D}
\bibinfo{author}{Knirck, S.} \emph{et~al.}
\newblock \bibinfo{title}{{Simulating MADMAX in 3D: requirements for dielectric
  axion haloscopes}}.
\newblock \emph{\bibinfo{journal}{J. Cosmol. Astropart. Phys.}}
  \textbf{\bibinfo{volume}{10}}, \bibinfo{pages}{034} (\bibinfo{year}{2021}).
\newblock \urlprefix\url{https://doi.org/10.1088/1475-7516/2021/10/034}.

\bibitem{article:QUAX_proposal}
\bibinfo{author}{Barbieri, R.} \emph{et~al.}
\newblock \bibinfo{title}{{Searching for galactic axions through magnetized
  media: The QUAX proposal}}.
\newblock \emph{\bibinfo{journal}{Phys. Dark Univ.}}
  \textbf{\bibinfo{volume}{15}}, \bibinfo{pages}{135} (\bibinfo{year}{2017}).
\newblock \urlprefix\url{http://dx.doi.org/10.1016/j.dark.2017.01.003}.

\bibitem{article:QUAX-ae_1}
\bibinfo{author}{Crescini, N.} \emph{et~al.}
\newblock \bibinfo{title}{{Operation of a ferromagnetic axion haloscope at $m_a
  = 58\,\mu$eV}}.
\newblock \emph{\bibinfo{journal}{Eur. Phys. J. C}}
  \textbf{\bibinfo{volume}{78}}, \bibinfo{pages}{703} (\bibinfo{year}{2018}).
\newblock \urlprefix\url{https://doi.org/10.1140/epjc/s10052-018-6163-8}.

\bibitem{article:QUAX-ae_2}
\bibinfo{author}{Crescini, N.} \emph{et~al.}
\newblock \bibinfo{title}{{Axion Search with a Quantum-Limited Ferromagnetic
  Haloscope}}.
\newblock \emph{\bibinfo{journal}{Phys. Rev. Lett.}}
  \textbf{\bibinfo{volume}{124}}, \bibinfo{pages}{171801}
  (\bibinfo{year}{2020}).
\newblock \urlprefix\url{https://doi.org/10.1103/PhysRevLett.124.171801}.

\bibitem{article:plasma}
\bibinfo{author}{Lawson, M.}, \bibinfo{author}{Millar, A.~J.},
  \bibinfo{author}{Pancaldi, M.}, \bibinfo{author}{Vitagliano, E.} \&
  \bibinfo{author}{Wilczek, F.}
\newblock \bibinfo{title}{{Tunable Axion Plasma Haloscopes}}.
\newblock \emph{\bibinfo{journal}{Phys. Rev. Lett.}}
  \textbf{\bibinfo{volume}{123}}, \bibinfo{pages}{141802}
  (\bibinfo{year}{2019}).
\newblock \urlprefix\url{https://doi.org/10.1103/PhysRevLett.123.141802}.

\bibitem{article:AQ-TI}
\bibinfo{author}{Marsh, D. J.~E.} \emph{et~al.}
\newblock \bibinfo{title}{{Proposal to Detect Dark Matter using Axionic
  Topological Antiferromagnets}}.
\newblock \emph{\bibinfo{journal}{Phys. Rev. Lett.}}
  \textbf{\bibinfo{volume}{123}}, \bibinfo{pages}{121601}
  (\bibinfo{year}{2019}).
\newblock \urlprefix\url{https://doi.org/10.1103/PhysRevLett.123.121601}.

\bibitem{article:TOORAD}
\bibinfo{author}{S.-Engel, J.} \emph{et~al.}
\newblock \bibinfo{title}{{Axion Quasiparticles for Axion Dark Matter
  Detection}}  (\bibinfo{year}{2020}).
\newblock \eprint{2102.05366}.

\bibitem{article:LC_circuit}
\bibinfo{author}{Sikivie, P.}, \bibinfo{author}{Sullivan, N.} \&
  \bibinfo{author}{Tanner, B.~D.}
\newblock \bibinfo{title}{{Proposal for Axion Dark Matter Detection Using an
  $LC$ Circuit}}.
\newblock \emph{\bibinfo{journal}{{Phys. Rev. Lett.}}}
  \textbf{\bibinfo{volume}{112}}, \bibinfo{pages}{131301}
  (\bibinfo{year}{2014}).
\newblock \urlprefix\url{https://doi.org/10.1103/PhysRevLett.112.131301}.

\bibitem{article:ABRACADABRA}
\bibinfo{author}{Kahn, Y.} \emph{et~al.}
\newblock \bibinfo{title}{{Broadband Resonant Approaches to Axion Dark Matter
  Detection}}.
\newblock \emph{\bibinfo{journal}{Phys. Rev. Lett.}}
  \textbf{\bibinfo{volume}{117}}, \bibinfo{pages}{141801}
  (\bibinfo{year}{2016}).
\newblock \urlprefix\url{https://doi.org/10.1103/PhysRevLett.117.141801}.

\bibitem{article:ABRACADABRA-10cm}
\bibinfo{author}{Ouellet, J.~L.} \emph{et~al.}
\newblock \bibinfo{title}{{First Results from ABRACADABRA-10\,cm: A Search for
  Sub-$\mu$eV, Axion Dark Matter}}.
\newblock \emph{\bibinfo{journal}{Phys. Rev. Lett.}}
  \textbf{\bibinfo{volume}{122}}, \bibinfo{pages}{121802}
  (\bibinfo{year}{2019}).
\newblock \urlprefix\url{https://doi.org/10.1103/PhysRevLett.122.121802}.

\bibitem{article:DM_Radio}
\bibinfo{author}{Chaudhuri, S.} \emph{et~al.}
\newblock \bibinfo{title}{{Radio for hidden-photon dark matter detection}}.
\newblock \emph{\bibinfo{journal}{Phys. Rev. D}} \textbf{\bibinfo{volume}{92}},
  \bibinfo{pages}{075012} (\bibinfo{year}{2015}).
\newblock \urlprefix\url{https://doi.org/10.1103/PhysRevD.92.075012}.

\bibitem{article:DM_Radio_design}
\bibinfo{author}{Silva-Feaver, M.} \emph{et~al.}
\newblock \bibinfo{title}{{Design Overview of the DM Radio Pathfinder
  Experiment}}  (\bibinfo{year}{2016}).
\newblock \eprint{1610.09344}.

\bibitem{article:SHAFT}
\bibinfo{author}{Gramolin, A.~V.} \emph{et~al.}
\newblock \bibinfo{title}{{Search for axion-like dark matter with
  ferromagnets}}.
\newblock \emph{\bibinfo{journal}{Nature Phys.}} \textbf{\bibinfo{volume}{17}},
  \bibinfo{pages}{79} (\bibinfo{year}{2021}).
\newblock \urlprefix\url{https://doi.org/10.1038/s41567-020-1006-6}.

\bibitem{article:BNL_Helioscope}
\bibinfo{author}{Lazarus, D.~M.} \emph{et~al.}
\newblock \bibinfo{title}{{Search for solar axions}}.
\newblock \emph{\bibinfo{journal}{Phys. Rev. Lett.}}
  \textbf{\bibinfo{volume}{69}}, \bibinfo{pages}{2333} (\bibinfo{year}{1992}).
\newblock \urlprefix\url{https://doi.org/10.1103/PhysRevLett.69.2333}.

\bibitem{article:Tokyo_Helioscope}
\bibinfo{author}{Ohta, R.} \emph{et~al.}
\newblock \bibinfo{title}{{Search for sub-electronvolt solar axions using
  coherent conversion of axions into photons in magnetic field and gas
  helium}}.
\newblock \emph{\bibinfo{journal}{Phys. Lett. B}}
  \textbf{\bibinfo{volume}{536}}, \bibinfo{pages}{18} (\bibinfo{year}{2002}).
\newblock \urlprefix\url{https://doi.org/10.1016/S0370-2693(02)01822-1}.

\bibitem{article:CAST}
\bibinfo{author}{Anastassopoulos, V.} \emph{et~al.}
\newblock \bibinfo{title}{{New CAST limit on the axion–photon interaction}}.
\newblock \emph{\bibinfo{journal}{Nature Phys.}} \textbf{\bibinfo{volume}{13}},
  \bibinfo{pages}{584} (\bibinfo{year}{2017}).
\newblock \urlprefix\url{https://doi.org/10.1038/nphys4109}.

\bibitem{article:IAXO1}
\bibinfo{author}{Irastorza, I.} \emph{et~al.}
\newblock \bibinfo{title}{{Towards a new generation axion helioscope}}.
\newblock \emph{\bibinfo{journal}{J. Cosmol. Astropart. Part.}}
  \textbf{\bibinfo{volume}{06}}, \bibinfo{pages}{013} (\bibinfo{year}{2011}).
\newblock \urlprefix\url{https://doi.org/10.1088/1475-7516/2011/06/013}.

\bibitem{article:IAXO2}
\bibinfo{author}{Armengaud, E.} \emph{et~al.}
\newblock \bibinfo{title}{{Conceptual design of the International Axion
  Observatory (IAXO)}}.
\newblock \emph{\bibinfo{journal}{J. Instrum.}} \textbf{\bibinfo{volume}{9}},
  \bibinfo{pages}{T05002} (\bibinfo{year}{2014}).
\newblock \urlprefix\url{https://doi.org/10.1088/1748-0221/9/05/T05002}.

\bibitem{article:ATLAS}
\bibinfo{author}{Kate, H. H. J.~T.}
\newblock \bibinfo{title}{{The ATLAS superconducting magnet system at the Large
  Hadron Collider}}.
\newblock \emph{\bibinfo{journal}{Physica C}} \textbf{\bibinfo{volume}{468}},
  \bibinfo{pages}{2137} (\bibinfo{year}{2008}).
\newblock \urlprefix\url{https://doi.org/10.1016/j.physc.2008.05.146}.

\bibitem{article:BFRT1}
\bibinfo{author}{Ruoso, G.} \emph{et~al.}
\newblock \bibinfo{title}{{Search for photon regeneration in a magnetic
  field}}.
\newblock \emph{\bibinfo{journal}{Z. Phys. C}} \textbf{\bibinfo{volume}{56}},
  \bibinfo{pages}{505} (\bibinfo{year}{1992}).
\newblock \urlprefix\url{https://doi.org/10.1007/BF01474722}.

\bibitem{article:BFRT2}
\bibinfo{author}{Cameron, R.} \emph{et~al.}
\newblock \bibinfo{title}{{Search for nearly massless, weakly coupled particles
  by optical techniques}}.
\newblock \emph{\bibinfo{journal}{Phys. Rev. D}} \textbf{\bibinfo{volume}{47}},
  \bibinfo{pages}{3707} (\bibinfo{year}{1993}).
\newblock \urlprefix\url{https://doi.org/10.1103/PhysRevD.47.3707}.

\bibitem{article:OSQAR1}
\bibinfo{author}{Pugnat, P.} \emph{et~al.}
\newblock \bibinfo{title}{{Results from the OSQAR photon-regeneration
  experiment: No light shining through a wall}}.
\newblock \emph{\bibinfo{journal}{Phys. Rev. D}} \textbf{\bibinfo{volume}{78}},
  \bibinfo{pages}{092003} (\bibinfo{year}{2008}).
\newblock \urlprefix\url{https://doi.org/10.1103/PhysRevD.78.092003}.

\bibitem{article:OSQAR3}
\bibinfo{author}{Ballou, R.} \emph{et~al.}
\newblock \bibinfo{title}{{New exclusion limits on scalar and pseudoscalar
  axionlike particles from light shining through a wall}}.
\newblock \emph{\bibinfo{journal}{Phys. Rev. D}} \textbf{\bibinfo{volume}{92}},
  \bibinfo{pages}{092002} (\bibinfo{year}{2015}).
\newblock \urlprefix\url{https://doi.org/10.1103/PhysRevD.92.092002}.

\bibitem{article:ALPS}
\bibinfo{author}{Ehret, K.} \emph{et~al.}
\newblock \bibinfo{title}{{New ALPS results on hidden-sector lightweights}}.
\newblock \emph{\bibinfo{journal}{Phys. Lett. B}}
  \textbf{\bibinfo{volume}{689}}, \bibinfo{pages}{149} (\bibinfo{year}{2010}).
\newblock \urlprefix\url{https://doi.org/10.1016/j.physletb.2010.04.066}.

\bibitem{article:enhanced_LSW}
\bibinfo{author}{Sikivie, P.}, \bibinfo{author}{Tanner, D.~B.} \&
  \bibinfo{author}{van Bibber, K.}
\newblock \bibinfo{title}{{Resonantly Enhanced Axion-Photon Regeneration}}.
\newblock \emph{\bibinfo{journal}{Phys. Rev. Lett.}}
  \textbf{\bibinfo{volume}{98}}, \bibinfo{pages}{172002}
  (\bibinfo{year}{2007}).
\newblock \urlprefix\url{https://doi.org/10.1103/PhysRevLett.98.172002}.

\bibitem{article:ALPSII}
\bibinfo{author}{Baehre, R.} \emph{et~al.}
\newblock \bibinfo{title}{{Any light particle search II — Technical Design
  Report}}.
\newblock \emph{\bibinfo{journal}{J. Instrum.}} \textbf{\bibinfo{volume}{8}},
  \bibinfo{pages}{T09001} (\bibinfo{year}{2013}).
\newblock \urlprefix\url{https://doi.org/10.1088/1748-0221/8/09/T09001}.

\bibitem{article:Hempelmann17}
\bibinfo{author}{Hempelmann, N.} \emph{et~al.}
\newblock \bibinfo{title}{{Phase locking the spin precession in a storage
  ring}}.
\newblock \emph{\bibinfo{journal}{Phys. Rev. Lett.}}
  \textbf{\bibinfo{volume}{119}}, \bibinfo{pages}{014801}
  (\bibinfo{year}{2017}).
\newblock \urlprefix\url{https://doi.org/10.1103/PhysRevLett.119.014801}.

\bibitem{article:Guidoboni16}
\bibinfo{author}{Guidoboni, G.}
\newblock \bibinfo{title}{{How to reach a Thousand-second in-plane Polarization
  Lifetime with 0.97\,GeV/c Deuterons in a storage ring}}.
\newblock \emph{\bibinfo{journal}{Phys. Rev. Lett.}}
  \textbf{\bibinfo{volume}{117}}, \bibinfo{pages}{054801}
  (\bibinfo{year}{2016}).
\newblock \urlprefix\url{https://doi.org/10.1103/PhysRevLett.117.054801}.

\bibitem{article:Anastassopoulos16}
\bibinfo{author}{Anastassopoulos, V.} \emph{et~al.}
\newblock \bibinfo{title}{{A storage ring experiment to detect a proton
  electric dipole moment}}.
\newblock \emph{\bibinfo{journal}{Rev. Sci. Instrum.}}
  \textbf{\bibinfo{volume}{87}}, \bibinfo{pages}{115116}
  (\bibinfo{year}{2016}).
\newblock \urlprefix\url{https://doi.org/10.1063/1.4967465}.

\bibitem{article:Metodiev15}
\bibinfo{author}{Metodiev, E.~M.} \emph{et~al.}
\newblock \bibinfo{title}{{Analytical benchmarks for precision particle
  tracking in electric and magnetic rings}}.
\newblock \emph{\bibinfo{journal}{Nucl. Instrum. Methods Phys. Res. A}}
  \textbf{\bibinfo{volume}{797}}, \bibinfo{pages}{311} (\bibinfo{year}{2015}).
\newblock \urlprefix\url{https://doi.org/10.1016/j.nima.2015.06.032}.

\bibitem{article:Metodiev14}
\bibinfo{author}{Metodiev, E.~M.}, \bibinfo{author}{Huang, K.~L.},
  \bibinfo{author}{Semertzidis, Y.~K.} \& \bibinfo{author}{Morse, W.~M.}
\newblock \bibinfo{title}{{Fringe electric fields of flat and cylindrical
  deflectors in electrostatic charged particle storage rings}}.
\newblock \emph{\bibinfo{journal}{Phys. Rev. ST Accel. Beams}}
  \textbf{\bibinfo{volume}{17}}, \bibinfo{pages}{074002}
  (\bibinfo{year}{2014}).
\newblock \urlprefix\url{https://doi.org/10.1103/PhysRevSTAB.17.074002}.

\bibitem{article:Morse13}
\bibinfo{author}{Morse, W.~M.}, \bibinfo{author}{Orlov, Y.~F.} \&
  \bibinfo{author}{Semertzidis, Y.~K.}
\newblock \bibinfo{title}{{rf Wien filter in an electric dipole moment storage
  ring: The “partially frozen spin” effect}}.
\newblock \emph{\bibinfo{journal}{Phys. Rev. ST Accel. Beams}}
  \textbf{\bibinfo{volume}{16}}, \bibinfo{pages}{114001}
  (\bibinfo{year}{2013}).
\newblock \urlprefix\url{https://doi.org/10.1103/PhysRevSTAB.16.114001}.

\bibitem{article:Brantjes12}
\bibinfo{author}{M.Brantjes, N.~P.}
\newblock \bibinfo{title}{{Correction systematic errors in high-sensitivity
  deuteron polarization measurements}}.
\newblock \emph{\bibinfo{journal}{Nucl. Instrum. Methods Phys. Res. A}}
  \textbf{\bibinfo{volume}{664}}, \bibinfo{pages}{49} (\bibinfo{year}{2012}).
\newblock \urlprefix\url{https://doi.org/10.1016/j.nima.2011.09.055}.

\bibitem{article:Bennett09}
\bibinfo{author}{Bennett, G.~W.} \emph{et~al.}
\newblock \bibinfo{title}{{Improved limit on the muon electric dipole momen}}.
\newblock \emph{\bibinfo{journal}{Phys. Rev. D}} \textbf{\bibinfo{volume}{80}},
  \bibinfo{pages}{052008} (\bibinfo{year}{2009}).
\newblock \urlprefix\url{https://doi.org/10.1103/PhysRevD.80.052008}.

\bibitem{article:Farley04}
\bibinfo{author}{Farley, F.~J.~M.} \emph{et~al.}
\newblock \bibinfo{title}{{New Method of Measuring Electric Dipole Moments in
  Storage Rings}}.
\newblock \emph{\bibinfo{journal}{Phys. Rev. Lett.}}
  \textbf{\bibinfo{volume}{93}}, \bibinfo{pages}{052001}
  (\bibinfo{year}{2004}).
\newblock \urlprefix\url{https://doi.org/10.1103/PhysRevLett.93.052001}.

\bibitem{article:Haciomeroglu19}
\bibinfo{author}{Haciomeroglu, S.} \& \bibinfo{author}{Semertzidis, Y.~K.}
\newblock \bibinfo{title}{{Hybrid ring design in the storage-ring proton EDM
  experiment}}.
\newblock \emph{\bibinfo{journal}{Phys. Rev. Accel. Beams}}
  \textbf{\bibinfo{volume}{22}}, \bibinfo{pages}{034001}
  (\bibinfo{year}{2019}).
\newblock \urlprefix\url{https://doi.org/10.1103/PhysRevAccelBeams.22.034001}.

\bibitem{article:Omarov20}
\bibinfo{author}{Omarov, Z.} \emph{et~al.}
\newblock \bibinfo{title}{{Comprehensive Symmetric-Hybrid ring design for pEDM
  experiment at below $10^{-29}\,e\cdot$cm}}  (\bibinfo{year}{2020}).
\newblock \eprint{2007.10332}.

\bibitem{article:ARIADNE_SC}
\bibinfo{author}{Fosbinder-Elkins, H.}, \bibinfo{author}{Kim, Y.} \emph{et~al.}
\newblock \bibinfo{title}{{A method for controlling the magnetic field near a
  superconducting boundary in the ARIADNE axion experiment}}.
\newblock \emph{\bibinfo{journal}{Quantum Sci. Technol.}}
  (\bibinfo{year}{2021}).
\newblock \urlprefix\url{https://doi.org/10.1088/2058-9565/abf1cc}.
\newblock \bibinfo{note}{Accepted}.

\bibitem{article:nEDM}
\bibinfo{author}{Abel, .} \emph{et~al.}
\newblock \bibinfo{title}{{Measurement of the Permanent Electric Dipole Moment
  of the Neutron}}.
\newblock \emph{\bibinfo{journal}{Phys. Rev. Lett.}}
  \textbf{\bibinfo{volume}{124}}, \bibinfo{pages}{081803}
  (\bibinfo{year}{2020}).
\newblock \urlprefix\url{https://doi.org/10.1103/PhysRevLett.124.081803}.

\end{thebibliography}

 \end{document}